\documentclass[a4paper]{article}
\usepackage[
  a4paper,
  left=3cm,
  right=3cm,
  top=3cm,
  bottom=3cm
]{geometry}
\usepackage{amsmath}
\usepackage{yhmath}
\usepackage{hyperref}
\usepackage{graphicx} % Required for inserting images
\usepackage{subcaption}    % 子图支持       % 图标题控制
\usepackage{float}         % 控制图像位置
\usepackage{amssymb,amsthm}
\usepackage{algorithm,algorithmicx,algpseudocode}
\usepackage{verbatim}
\usepackage[shortlabels]{enumitem}
\usepackage{parskip}
\usepackage{xcolor}
\theoremstyle{plain}

\newtheorem{theorem}{Theorem}[section]
\newtheorem{proposition}[theorem]{Proposition}
\newtheorem{lemma}[theorem]{Lemma}
\newtheorem{corollary}[theorem]{Corollary}

\theoremstyle{definition}
\newtheorem{definition}[theorem]{Definition}
\theoremstyle{remark}
\newtheorem{remark}[theorem]{Remark}

\usepackage{anyfontsize}
\newcommand{\R}{\mathbb{R}}
\newcommand{\N}{\mathbb{N}}

\newcommand{\Rm}{\mathbb R}

\newcommand{\fa}{{\mathfrak a}}
\newcommand{\py}{\partial_y}
\newcommand{\epm}{\epsilon_m}
\newcommand{\phm}{\phi_m}
\newcommand{\psm}{\psi_m}
\newcommand{\xim}{\xi_m}

\newcommand{\sgn}{\mathrm{sign}}

\newcommand{\iter}{\mathrm{i}}
\newcounter{experiment}
\newcommand{\rev}[1]{{\color{black}#1}}
\DeclareSymbolFont{yhlargesymbols}{OMX}{yhex}{m}{n}
\DeclareMathAccent{\reallywidehat}{\mathord}{yhlargesymbols}{"62}

\title{Inverse scattering for waveguides in topological insulators}

\author{Guillaume Bal\thanks{Departments of Statistics and Mathematics and CCAM, University of Chicago,  guillaumebal@uchicago.edu}, Xixian Wang\thanks{Division of Mathematical Sciences, School of Physical and Mathematical Sciences, Nanyang Technological University, Singapore. xixian001@e.ntu.edu.sg}, Zhongjian Wang\thanks{Division of Mathematical Sciences, School of Physical and Mathematical Sciences, Nanyang Technological University, Singapore. zhongjian.wang@ntu.edu.sg (Corresponding)}}

\begin{document}

\maketitle

\begin{abstract}
This paper concerns the inverse scattering problem of a topologically non-trivial waveguide separating two-dimensional topological insulators. We consider the specific model of a Dirac system. For scalar short-range perturbations, we prove linearized uniqueness and stability and obtain local uniqueness in a finite-dimensional setting under a smallness constraint. For general Hermitian perturbations, the scattering data are invariant under a natural gauge transformation; at the linearized level, the data determine the potential modulo precisely this gauge. We then solve the problem numerically by means of a standard adjoint method and illustrate our theoretical findings with several numerical simulations.
\end{abstract}
\textbf{keywords:}{topological insulators, Dirac operator, asymmetric transport, waveguide,  inverse scattering}

\section{Introduction}

This paper concerns the {inverse scattering theory} of one-dimensional interfaces separating two-dimensional insulating bulks {in the context} of topological insulators; see, e.g., \cite{BH, delplace, PS, sato, Volovik, Witten} for applications in many areas of condensed matter physics, photonics, and the geophysical sciences. A characteristic feature of the waveguide generated near the separating interface is a quantized asymmetric transport that affords a topological origin, making it robust to perturbations. 

As a concrete example of such an interface, we consider a two-dimensional Dirac system modeled by the following Hamiltonian:
\begin{equation}
    \label{eq:Dirac0}
  { H= D\cdot\sigma + m(y)\sigma_3 + V(x,y), \quad m(y)=y,}
\end{equation}
where $(x,y)\in{\mathbb R}^2$ are Cartesian coordinates, $D\cdot\sigma=-i\partial_x\sigma_1-i\partial_y\sigma_2$ for $\sigma_{1,2,3}$ the standard Pauli matrices, and where $V(x,y)$ is a $2\times2$ Hermitian-valued smooth function with compact support to simplify the presentation. 
Here, $m(y)=y$ is a mass term modeling the transition between the north insulating phase $m(y)>0$ for $y\geq1$ to the south insulating phase $m(y)<0$ for $y\leq -1$. This generates a waveguide in the vicinity of $y\approx0$. The salient feature of this two-dimensional model is a combination of transport along the $x$ axis near $y\approx0$ and confinement in the transverse $y-$variable. The above model serves as a prototypical example of a topologically non-trivial interface separating insulating bulks in different topological phases. See \cite{2} for results on the topological classification of Dirac operators and, e.g., \cite{bal2025topological} and references there for generalizations to other partial differential models. For the spectral analysis below, we use the unitarily equivalent
representation obtained by conjugating \eqref{eq:Dirac0} with
$Q=\frac{1}{\sqrt{2}}(\sigma_1+\sigma_3)$ and write $V$ for $QVQ$, as detailed in
Section~\ref{sec:prelim}. {We also consider the related, but topologically trivial, Klein Gordon operator with partial harmonic confinement $H_{KG}=-\Delta+y^2+V(x,y)$, also in two space dimensions.}

In the absence of the perturbation (i.e., $V=0$), the spectral decomposition of the corresponding $H_0$ gives rise to a number of propagating solutions $(H_0-E)\psi=0$, of the form $\psi(x,y)=e^{ix\xi(E)}\phi(y;E)$. Such `confined plane waves' are modified in the presence of the perturbation $V$, {leading} to a spectral decomposition of $H=H_0+V$ with generalized plane waves of the form $\psi^V(x,y)$. The {forward scattering} theory {leading to} the existence of such plane waves is {developed} in \cite{chen2025scattering}. %, following a limiting absorption theory developed in \cite{ASNSP_1975_4_2_2_151_0}.  
These modified plane waves $\psi^V$ away from the support of $V$ may then be decomposed over the unperturbed plane waves, giving rise to scattering coefficients, also shown to be well defined for the above Dirac model in \cite{chen2025scattering}. The details of that construction are recalled in section \ref{sec:prelim}
.
The present work considers the {\em inverse scattering} problem, namely the reconstruction of $V(x,y)$ from such scattering data $S(E)$.

\medskip

Inverse scattering problems have been thoroughly studied in the setting of operators of the form $H=H_0+V$ where $H_0$ is a constant-coefficient operator, typically a Laplacian or a Maxwell operator, for many practical applications \cite{colton-kress-98,isakov-2010}, and for fairly general one-dimensional systems in \cite{beals1984scattering}. There is significantly less work on cases where some of the physical dimensions are confined, as in the case of waveguides; see, e.g., \cite{arens2011direct,bourgeois2023scattering,isozaki2010forward} for references on inverse scattering problems for second-order equations with geometric confinement. 
We are not aware of any work treating the setting of topological waveguides, which is the main objective of this paper. As scattering theories for general topological waveguides are not widely available beyond the cases treated in \cite{chen2025scattering}, we consider here the inverse scattering problem specifically for the Dirac operator in \eqref{eq:Dirac0}.

%{Compared with standard scalar waveguides, a key feature of this model is the asymmetric lowest edge branch: the mode \((0,-)\) is present while the opposite branch \((0,+)\) is absent. As shown in Section~\ref{sec:main}, this asymmetric structure affects the accessibility of longitudinal Fourier frequencies in the inverse reconstruction.} %%% This is much too detailed without context.
{The interface} asymmetry is a manifestation of the topologically protected transport along the \(x\)-axis in the negative direction (towards \(-\infty\)), independently of the choice of Hermitian-valued perturbation \(V\)~\cite{2,BH}; {see Section \ref{sec:prelim} and Section \ref{sec:main} for additional detail}. This {asymmetry} may be {interpreted} as a topological obstruction to Anderson localization \cite{1,bal2023mathbb,PS}. \rev{A separate obstruction is intrinsic to the Dirac inverse problem, even for smooth compactly supported perturbations. Indeed, scalar conjugation by $e^{i\theta}$ leaves the scattering data unchanged and transforms the potential according to
\begin{equation*}
    V\longmapsto V+(\partial_x\theta)\sigma_3-(\partial_y\theta)\sigma_2,
    \qquad \theta\in C_c^\infty(\mathbb R^2;\mathbb R).
\end{equation*}
Consequently, a general Hermitian potential cannot be uniquely reconstructed from scattering data. We show instead that the linearized data determine the gauge-invariant components, while scalar potentials can be reconstructed uniquely and stably.
More precisely, {first considering} the linearized inverse scattering problem (linearized about $V=0$), {we} prove an explicit Fourier-domain reconstruction and stability estimate for scalar potentials. We next show that a scalar potential represented in a finite-dimensional space can be uniquely and stably reconstructed from scattering data provided that it is sufficiently small. For a general Hermitian potential $V=\sum_{j=0}^3v_j\sigma_j$, the linearized data determine $v_0$, $v_1$, and the gauge-invariant combination $\partial_xv_2+\partial_yv_3$.  These results are {detailed} in Section \ref{sec:main} and Appendix~\ref{app:full}.}

Beyond a theoretical analysis of the reconstruction of $V$ in the linearized setting and in the finite-dimensional setting, we also provide an algorithm to solve the inverse scattering problem. Based on the algorithm to solve the forward scattering problem developed in \cite{bal2023asymmetric}, we use a standard adjoint method \cite{chavent2009nonlinear,vogel2002computational} that enables efficient gradient computation with respect to a discretized potential expansion. The numerical implementation does not impose the smallness condition used
in the local theoretical result.

The rest of the paper is structured as follows. Section \ref{sec:prelim} recalls the necessary information on the spectral decomposition of the Dirac operator and on the scattering data generated by the perturbation $V(x,y)$. Our main results are described in Section \ref{sec:main}.
Section \ref{sec:adjoint} then presents the adjoint method, an iterative procedure to solve the inverse scattering problem. Numerical simulations illustrate the procedure in Section \ref{sec:num}. Concluding remarks are presented in Section \ref{sec:conclu} while some lemmas and proofs are postponed to the appendix. {The inverse scattering problem for the harmonic confinement $H_{KG}$ is also detailed in an appendix.}

\section{Forward and inverse scattering problem}\label{sec:prelim}
We first describe the scattering data in the two-dimensional Dirac model, mostly following \cite{1,bal2023asymmetric}, and {next} present the linearized inverse problem and the small-potential inverse problem.
\paragraph{Spectral decomposition.} We start with the {unperturbed} two-dimensional Dirac operator,
  \begin{align}
    \label{eq:Dirac}
    H_0=D_x \sigma_3 - D_y \sigma_2 + m(y) \sigma_1  = \begin{pmatrix}
    D_x & m(y)+iD_y  \\ m(y)-iD_y & -D_x
    \end{pmatrix},
\end{align}
where $\sigma_{1,2,3}$ are the standard Pauli matrices, which, along with the identity matrix $\sigma_0=I_2$, form a basis of $2\times2$ Hermitian matrices. $D_a=-i\partial_a$ for $a=x,y$, and $m(y)=y$ denotes a linear domain wall. The above Dirac operator is unitarily equivalent to \eqref{eq:Dirac0} by a unitary transformation mapping $H$ to $QHQ$ with $Q=\frac{1}{\sqrt2}(\sigma_1+\sigma_3)$ (we still call $H$ the resulting Hamiltonian in \eqref{eq:Dirac}).

Let $\fa=\py+y$ and define the Hermite functions,
\begin{equation}\label{eq:Hermite functions}
    \varphi_n(y)=a_n(\fa^*)^n\varphi_0(y),
\end{equation}
where $\varphi_0(y)=\pi^{-\frac{1}{4}}e^{-\frac{1}{2}y^2}$ and $a_n$ is a normalizing factor. These functions form an orthonormal basis of $L^2(\R_y)$ and satisfy,
\begin{equation*}
    \fa^*\fa \varphi_n=2n \varphi_n, \quad \fa \varphi_n=\sqrt{2n}\varphi_{n-1}, \quad \fa^*\varphi_n=\sqrt{2(n+1)}\varphi_{n+1}.
\end{equation*}
We define the countable index set $M$ as the union of the following pairs of indices $m$. 
For $n\in\N^+$, let $m=(n,\epsilon_m)$ with $\epsilon_m=\pm1$, while for $n=0$, we only include $m=(0,-1)$. {Thus an index \(m\in M\) consists of a transverse mode level \(n\) and a sign
\(\epsilon_m\).
%For \(n\geq 1\), the two signs correspond to the two branches
%\(\xi_m(E)=\epsilon_m(E^2-2n)^{1/2}\). 
%%%% This notation is not defined yet. Cannot be used. 
The exceptional index \((0,-1)\)
corresponds to the asymmetric lowest edge branch; what may be interpreted as an opposite branch
\((0,+1)\) is absent in this Dirac domain-wall model.} For $m=(n,\epsilon_m)$,
we write $|m|=n$. For simplicity, we also write $n\pm$ to denote $(n,\pm1)$.

{For each fixed energy $E$, the homogeneous equation
$(H_0-E)\psi=0$
is solved by separating the longitudinal variable $x$ from the transverse
variable $y$, and the normalized solutions are given by }
\begin{equation*}
    \psm(x,y;E)=e^{i\xim(E) x}\phm(y;E), \quad m \in M,
\end{equation*}
where $\xim(E)=\epm(E^2-2n)^{\frac{1}{2}}$, $\sqrt{-1}=i$,  $ \xi_{(0,-1)}=-E $, $\phi_{(0,-1)}=(0,\varphi_0)^T$ when $m=(0,-1)$ and for other $m=(n,\epsilon_m)\in M(E)$,
\begin{equation}\label{eq:eigenfunctions y;E}
    \begin{aligned}    
    \phm(y;E)&=c_m\begin{pmatrix}
        \fa\varphi_n(y)\\ (E-\xim)\varphi_n(y)
    \end{pmatrix}=c_m\begin{pmatrix}
        \sqrt{2n}\varphi_{n-1}(y)\\(E-\xim)\varphi_n(y)
    \end{pmatrix}, \quad 
    c_m=1/\sqrt{2n+|E-\xi_m|^2}.    
\end{aligned}
\end{equation}{The spinor $\phi_m(y;E)$ describes the
transverse profile associated with the mode level $n$, while the factor $e^{i\xi_m(E)x}$ describes the behavior in
the longitudinal direction.}
 When $\xi_m$ is real-valued, $\psi_m$ does not vanish as $x\to \pm \infty$ and is referred to as a propagating mode and the longitudinal group velocity is
$\xi_m(E)/E$. Incoming and outgoing channels are classified
by the sign of this velocity. For $E>0$, this agrees with
the branch sign $\epsilon_m$; the exceptional mode has
velocity $-1$ for both signs of $E$. When $\xi_m$ is purely imaginary, $\psi_m$ is an evanescent mode. For each $E\in \R$, we further define the energy-dependent subset of \(M\) consisting of propagating modes,
\begin{align*}
    M(E):=\{\,m=(n,\epsilon_m)\in M \mid  E^2-2n>0\,\}.
\end{align*}

A known feature of the eigen-system of $H$ is the following stability under perturbation \cite{chen2025scattering}. %Define $H=H_0+V$.
Consider a Hermitian-valued perturbation $V(x,y)$ and the perturbed system,
\begin{equation}\label{perturbed dirac}
     (H-E)\psi=0,\quad H=H_0+V.
\end{equation}
Let $\psi_{\mathrm{in}}$ be a generalized eigenfunction of the unperturbed operator $H_0$, satisfying
$(H_0 - E)\psi_{\mathrm{in}} = 0$ for some fixed $E$. We look for the outgoing solutions of
\begin{equation}
    \label{eq:outgoing solution}
    (H-E)\psi_{out}=-V\psi_{in}.
\end{equation}
Then the (generalized) eigenfunction $\psi=\psi_{in}+\psi_{out}$ satisfies \eqref{perturbed dirac}. We consider the following integral formulation of $\psi_{out}$ under the outgoing Green's function of {$H_0-E$},
\begin{equation*}
    \psi_{out}(x,y;E)
    = \int G_{out}(x,y;x_0,y_0;E)
      \rho(x_0,y_0;E)dx_0dy_0,
\end{equation*}
where $\rho$ is the source density associated with $\psi_{out}$.
Defining
\begin{align*}
    \theta_n(E)=\begin{cases}
        i\sgn(E)\sqrt{E^2-2n}, & \qquad E^2>2n,\\
          -\sqrt{2n-E^2}, &\qquad E^2<2n.
    \end{cases}
\end{align*}
The outgoing Green's function $G_{out}$ has the following explicit form \cite{bal2023asymmetric},
\begin{align*}
    G_{out}=\begin{pmatrix}
        (D_x+E)G_{out,+} & \fa G_{out,-}\\
        \fa^*G_{out,+} & (-D_x+E)G_{out,-}
    \end{pmatrix},
\end{align*}
%where
\begin{equation}\label{eq:G_out + -}
    \begin{aligned}
    &G_{out,-}(x,y;x_0,y_0;E)=\sum_{n= 0}^{\infty}\frac{-1}{2\theta_n(E)}e^{\theta_n(E)|x-x_0|}\varphi_n(y)\varphi_n(y_0),\\
    &G_{out,+}(x,y;x_0,y_0;E)=\sum_{n= 1}^{\infty}\frac{-1}{2\theta_{n}(E)}e^{\theta_{n}(E)|x-x_0|}\varphi_{n-1}(y)\varphi_{n-1}(y_0) . %,
\end{aligned}
\end{equation}
%and $\theta_n(E)=i\sqrt{E^2-2n}$.

\paragraph{Scattering matrix.}
 Suppose $V$ bounded and decaying rapidly in the $x$-direction such that 
 \begin{align}\label{eq:V decay fast in x}
     (1+x^2)^{\frac{h}{2}}|V(x,y)|\le C,\quad \forall (x,y)\in \R^2,
 \end{align} for some constant $C$ and $h>1$. By Proposition~6.4 in \cite{chen2025scattering}, the solution $\psi$ of \eqref{perturbed dirac} satisfies:
\begin{align}
	\psi(x,y)=\sum_{m\in M} \alpha_m(x)\phi_m(y) \approx\sum_{m\in M(E)} \beta_m(\pm)\tilde{\psi}_m(x,y),\label{decomp inf}
\end{align}
where {$\tilde{\psi}_m(x,y):=\sqrt{|\frac{E}{\xi_m}|}\psi_m(x,y)$}, and where $a\approx b$ means that $\|a(x,y)-b(x,y)\|_{L^2(\R_y)}$ converges to $0$ uniformly as $x\rightarrow \pm \infty$. {The coefficients $ \beta_m(\pm)$  denote the corresponding asymptotic modal amplitudes in the 
limits $x\to\pm\infty$. Since evanescent 
modes decay exponentially away from the support of $V$, only propagating 
modes $m\in M(E)$ appear in this far-field asymptotic expansion.} 

Based on the above asymptotic decomposition, the generalized eigenfunctions can be characterized by a linear map between the incoming and outgoing coefficients. More precisely, the incoming conditions are given by coefficients of the right-traveling modes $\tilde{\psi}_m$ ($\epsilon_m>0$) at the left side and the left-traveling modes $\tilde{\psi}_m$ ($\epsilon_m<0$) at the right side, namely  $\beta_-(+)$ and $\beta_+(-)$. The outgoing solution by \eqref{eq:outgoing solution} is given by the coefficients of the left-traveling modes $\tilde{\psi}_m$, ($\epsilon_m<0$), at left side and the right-traveling modes $\tilde{\psi}_m$ ($\epsilon_m>0$) at right side, namely  $\beta_-(-)$ and $\beta_+(+)$. Then, the scattering matrix $S$ for $V$ can be defined as,
\begin{align}\label{def:Scattering Matrix}
    \begin{pmatrix}
    \beta_{-}(-)\\
    \beta_{+}(+)
    \end{pmatrix}=
    S
    \begin{pmatrix}
    \beta_{-}(+)\\
    \beta_{+}(-)
    \end{pmatrix}.
\end{align}
Under such a normalization, $S$ is unitary \cite{chen2025scattering} and when $V=0$, $S$ is the identity matrix.

 When $V$ is compactly supported in the $x$-direction on the interval $[x_L,x_R]$, the transmission reflection (TR) matrix, which also accounts for the evanescent modes, is defined by,
\begin{align}\label{def:TR Matrix}
    \begin{pmatrix}
    \alpha_{-}(x_L)\\
    \alpha_{+}(x_R)
    \end{pmatrix}=
    T
    \begin{pmatrix}
    \alpha_{-}(x_R)\\
    \alpha_{+}(x_L)
    \end{pmatrix},
\end{align}
{where $\alpha_+(x)$ denotes the vector of modal coefficients
$(\alpha_m(x))_{\epsilon_m=+1}$, and $\alpha_-(x)$ denotes
$(\alpha_m(x))_{\epsilon_m=-1}$.}
For each $ p=(q,\epsilon_p)\in M$, denote the incoming wave $\psi_{in}^p$ by
%$\psi_{in}^p(x,y):=  e^{i\xi_p x}\phi_p(y)$,
\begin{equation}\label{eq:psi_in k}
    \psi_{in}^p(x,y)=
    \begin{cases}
       e^{i\xi_p (x-x_L)}\phi_p(y), \quad \epsilon_p>0,\\
       e^{i\xi_p (x-x_R)}\phi_p(y), \quad \epsilon_p<0,
    \end{cases}
\end{equation}
and let $\psi_{out}^p$ be the corresponding outgoing solution, i.e. 
 \begin{equation}\label{eqn:psi_dcomp}
     (H-E)\psi_{out}^p=-V\psi_{in}^p.
 \end{equation}
Denote the wave $\psi^p:=\psi_{in}^p+\psi_{out}^p=\sum_{m\in M}\alpha_{m}^p(x)\phi_{m}(y)$.
Then, by Cramer's rule,
\begin{align}\label{eq:alpha ^p _m}
\alpha^p_m(x)=\langle\vartheta_m(y),\psi^p(x,y) \rangle_y,
\end{align}
%
%Here,
\begin{align*}
    \vartheta_{n,-}(y):=\frac{1}{1-|P_n|^2}\big (\phi_{n-}(y)-P_n\phi_{n+}(y)\big),\\
     \vartheta_{n,+}(y):=\frac{1}{1-|P_n|^2}\big (\phi_{n+}(y)-\overline{P_n}\phi_{n-}(y)\big),
\end{align*}
where we denote $\phi_{0+}=0$ and $P_n=\langle \phi_{n+}(y),\phi_{n-}(y)\rangle_y $ for $n \in \N $. The TR matrix $T$ is then given by $T_{m,p}=\alpha_m^p(x_L)$ for $\epsilon_m=-1$ and $T_{m,p}=\alpha_m^p(x_R)$ for $\epsilon_m=1$.

The TR matrices are used to compute the complete eigen-systems of \eqref{perturbed dirac}, which include the evanescent modes, and admit merging operation between adjacent intervals. A merging-based fast numerical algorithm~\cite{bal2023asymmetric} can then be applied to compute such eigenfunctions and is described in detail in Appendix~\ref{app:TRalg} for completeness.
\paragraph{Inverse Scattering Problem.} 
%Suppose we observe a TR matrix $T^{ob}$, while the underlying potential $V$ is unknown.  
The goal of the inverse scattering problem is to reconstruct $V$ from an observed scattering matrix $S^{ob}$ or, more generally, the TR matrix $T^{ob}$. More precisely, we seek a potential $V_0$ that solves
\begin{equation}\label{eq:optim}
    V_0  =\arg \min \Pi^S (V):=\arg\min \|S^{ob}-\mathcal{L}(V)\|,
\end{equation}
where $\|\cdot\|$ denotes some appropriate norm  in the observation space and $\mathcal{L}(V)$ is the scattering matrix of the perturbed scattering problem in \eqref{def:Scattering Matrix}.

The formulation~\eqref{eq:optim} raises the question of whether the observed
data identify a unique potential. For general Hermitian potentials, there
is an obstruction: a compactly supported change of phase produces a
different potential with exactly the same scattering data. The following
proposition records this symmetry of the full scattering map.
\rev{\begin{proposition}\label{prop:compact-gauge}
Let $\theta\in C_c^\infty(\mathbb R^2;\mathbb R)$ and set
\begin{equation}\label{eq:gauge-potential}
    V_\theta=(\partial_x\theta)\sigma_3-(\partial_y\theta)\sigma_2.
\end{equation}
Then $H_0+V+V_\theta=e^{-i\theta}(H_0+V)e^{i\theta}$, and $H_0+V$ and
$H_0+V+V_\theta$ have the same scattering data. In particular, neither the
nonlinear scattering map nor its linearization is injective on general
compactly supported Hermitian potentials.
\end{proposition}
\begin{proof}
Since $e^{-i\theta}D_ae^{i\theta}=D_a+\partial_a\theta$ for $a=x,y$ and
the scalar multiplier $e^{i\theta}$ commutes with $V$, the conjugation identity
follows from~\eqref{eq:Dirac}. Moreover, $e^{i\theta}=1$ outside
$\operatorname{supp}\theta$. Conjugation therefore leaves every incoming and
outgoing modal coefficient unchanged. Differentiating the identity along
$t\mapsto V+tV_\theta$ at $t=0$ shows that $V_\theta$ belongs to the kernel of
the linearized scattering map.
\end{proof}

For $V=\sum_{j=0}^3v_j\sigma_j$, the transformation in
Proposition~\ref{prop:compact-gauge} leaves
\[
    v_0,\qquad v_1,\qquad \partial_xv_2+\partial_yv_3
\]
unchanged. This suggests formulating the general Hermitian inverse problem
in terms of these gauge-invariant quantities. Their recovery from the full
linearized data is established in Appendix~\ref{app:full} in the Hermite
setting. For scalar potentials $V=v_0\sigma_0$, this particular ambiguity
is absent, since a nonzero $V_\theta$ cannot be the difference of two scalar
potentials. Section~\ref{sec:main} establishes uniqueness and stability in
this scalar class.}

\paragraph{Small potential linearization.}
To prepare both reconstruction analyses, we next derive the linearized
scattering map at $V=0$, retaining a general Hermitian perturbation $V$
throughout this derivation.

For small perturbation $V$,  the source density $\rho$ of $\psi_{out}$ can be formally approximated as
\begin{equation*}
    \rho = - (I + V\mathcal{G})^{-1} V \psi_{in}
          = -V\psi_{in} + \mathcal{O}(|V|^2),
\end{equation*}
where $\mathcal{G} = (H_0 - E-i0)^{-1}$ denotes the outgoing resolvent operator of $H_0$.  We denote the linearized solution as $
    \rho^{\mathrm{lin}}(x,y;E) = -V(x,y)\psi_{in}(x,y;E)$. 
Substituting this linearization into~\eqref{eq:outgoing solution} yields the linearized scattering problem,
\begin{equation}\label{eq:linearized problem}
    (H_0 - E)\psi_{out}^{\mathrm{lin}} = -V\psi_{in},
\end{equation}
whose solution can be written explicitly in terms of the outgoing Green's function as
\begin{equation}\label{eq:linearized solution}
\psi_{out}^{\mathrm{lin}}(x,y;E)
    = -\int G_{out}(x,y;x_0,y_0;E)
      V(x_0,y_0)\psi_{in}(x_0,y_0;E)dx_0dy_0.
\end{equation}
This provides a starting point to analyze the regularity of the non-linear scattering problem $\mathcal{L}$.
To this end, we first explicitly define the following space to ensure boundedness of $\mathcal{G}$. 

\paragraph{Small potential nonlinear inverse problem.}

We recall the following metric spaces \cite{chen2025scattering}.

\begin{definition}
Let $\langle x\rangle=\sqrt{1+x^2}$. For $s\in \mathbb{R}$ and $p\ge0$, we define the weighted Sobolev spaces $L_s^2$ and $H_s^p$ as the completion of $C_s^\infty(\mathbb{R}^2)$ under the norms
\begin{align*}  &\|u\|_{L^2_s}:=\Big(\int_{\mathbb{R}^2}\langle x\rangle^{2s}|u(x,y)|^2dxdy\Big)^{1/2},\\
    &\|u\|_{H^p_s}:=\Big(\int_{\mathbb{R}^2}\langle x\rangle^{2s}\big(\langle y \rangle^{2p}|u(x,y)|^2+\sum_{|\alpha|=p}|D^\alpha u (x,y) |^2\big)dxdy\Big)^{1/2}.
\end{align*}
%where $\langle x\rangle=\sqrt{1+x^2}$. 
\end{definition}
By Theorem~2.7 and Propositions~6.1--6.2 of \cite{chen2025scattering}, 
$
\mathcal{G}\in \mathcal{B}(L_1^2,H_{-1}^1).
$
Moreover, for a potential $V$ satisfying \eqref{eq:V decay fast in x}, 
viewed as a multiplication operator, we have
$
V\in \mathcal{B}(H_{-1}^1,L_1^2).
$
It follows that
$
\mathcal{G}V\in \mathcal{B}(H_{-1}^1,H_{-1}^1),
$
and the operator norm admits the bound
\begin{align*}
    \|\mathcal{G}V\|_{\mathcal{B}(H_{-1}^1,H_{-1}^1)}
    \le 
    \|\mathcal{G}\|_{\mathcal{B}(L_1^2,H_{-1}^1)}
    \|V\|_{\mathcal{B}(H_{-1}^1,L_1^2)}.
\end{align*}

In particular, for $\|V\|_{\mathcal{B}(H_{-1}^1,L_1^2)}$ sufficiently small,
we have $\|\mathcal{G}V\|_{\mathcal{B}(H_{-1}^1,H_{-1}^1)}<1$
%\begin{align*}
%\|\mathcal{G}V\|_{\mathcal{B}(H_{-1}^1,H_{-1}^1)}<1,
%\end{align*}
so that the inverse $(I+\mathcal{G}V)^{-1}$ exists on $H_{-1}^1$ and satisfies
the uniform bound
\begin{align*}
\|(I+\mathcal{G}V)^{-1}\|_{\mathcal{B}(H_{-1}^1,H_{-1}^1)}
\le \frac{1}{1-\|\mathcal{G}V\|_{\mathcal{B}(H_{-1}^1,H_{-1}^1)}}.
\end{align*}

Consequently, since
$\psi_{\mathrm{out}}-\psi_{\mathrm{out}}^{\mathrm{lin}}
=
(I+\mathcal{G}V)^{-1}(\mathcal{G}V)^2\psi_{\mathrm{in}}$,
we obtain the estimate
\begin{align*}
\|\psi_{\mathrm{out}}-\psi_{\mathrm{out}}^{\mathrm{lin}}\|_{H_{-1}^1}
\le 
\|(I+\mathcal{G}V)^{-1}\|_{\mathcal{B}(H_{-1}^1,H_{-1}^1)}
\|\mathcal{G}V\|_{\mathcal{B}(H_{-1}^1,H_{-1}^1)}^2
\|\psi_{\mathrm{in}}\|_{H_{-1}^1}.
\end{align*}
Dividing both sides by $\|V\|_{\mathcal{B}(H_{-1}^1,L_1^2)}$ and letting
$\|V\|_{\mathcal{B}(H_{-1}^1,L_1^2)}\to 0$, we conclude that
\begin{align}\label{eq:frechet deri}
\frac{\|\psi_{\mathrm{out}}-\psi_{\mathrm{out}}^{\mathrm{lin}}\|_{H_{-1}^1}}
    {\|V\|_{\mathcal{B}(H_{-1}^1,L_1^2)}}
    \leq C \|V\|_{\mathcal{B}(H_{-1}^1,L_1^2)}.
\end{align}
%Same for (20) below.
This shows that $\psi_{\mathrm{out}}^{\mathrm{lin}}$ provides the first-order
(Fr\'echet) approximation of $\psi_{\mathrm{out}}$ with respect to the potential
$V$. 

Assuming $V$ bounded and compactly supported in $[x_L,x_R]\times\Rm$, we consider %the decomposition
\begin{align*}
\psi_{out}(x,y;E;V)&=\sum_{m\in M}\beta_m(x;E;V)\tilde{\psi}_m(x,y;E),\\
\psi_{out}^{\mathrm{lin}}(x,y;E;V)&=\sum_{m\in M}\beta_m^{\mathrm{lin}}(x;E;V)\tilde{\psi}_m(x,y;E).
\end{align*}
The coefficients 
$\beta_m(x,E;V)$ and  $\beta_m^{\mathrm{lin}}(x,E;V)$ are {then} constant 
in $(-\infty,x_L]$ and $[x_R,\infty)$. 
Define
\begin{equation*}
    \hat{S}^{(\cdot)}(E;V) := S^{(\cdot)}(E;V) - I,
\end{equation*}
where $(\cdot)$ stands for either the full or linearized case. 
For each $p \in M(E)$, we impose the incoming condition $\psi_{in}=\widetilde\psi_p$. Then, for all $m \in M(E)$, the associated scattering coefficients satisfy
\begin{equation}\label{eq:scattering compact form}
    \hat{S}^{(\cdot)}_{m,p}(E;V) =
    \begin{cases}
        \beta_m^{(\cdot)}(x_L;E;V), & \epsilon_m < 0, \\[4pt]
        \beta_m^{(\cdot)}(x_R;E;V), & \epsilon_m > 0.
    \end{cases}
\end{equation}

By~\eqref{eq:frechet deri} and {since} the propagating coefficients of $\psi_{out}$ and $\psi_{out}^{\mathrm{lin}}$ are constant outside $[x_L,x_R]$, we have
\begin{align}\label{eq:linearized scatter 1 order app scatter}
     %\lim _{\|V\|_{\mathcal{B}(H^1_{-1},L^1_1)}\rightarrow 0} 
     \frac{|\hat{S}_{m,p}(E;V)-\hat{S}^{\mathrm{lin}}_{m,p}(E;V)|}{\|V\|_{\mathcal{B}(H^1_{-1},L^2_1)}}\leq C \|V\|_{\mathcal{B}(H_{-1}^1,L_1^2)}. 
\end{align}
{\begin{remark}
The forward scattering theory applies under the short-range
assumption \eqref{eq:V decay fast in x}. To simplify the presentation, we now assume the perturbation to be compactly supported. Extending the arguments to
the non-compact setting requires standard technicalities beyond the scope of this work. We simply note that the Fourier-domain reconstruction formulas obtained below remain valid for short-range
potentials.
\end{remark}
}
\paragraph{Explicit linearized scattering data.}

By~\eqref{eq:linearized solution} and~\eqref{eq:G_out + -}, $\hat{S}_{m,p}^\mathrm{lin}$ admits the explicit expression:
\begin{align}\label{eq:linearscat}
\hat{S}^{\mathrm{lin}}_{m,p}(E;V)
=-\frac{i|E|}
{\sqrt{|\xi_m(E)|\,|\xi_p(E)|}}\iint 
e^{-i\xi_{m}(E)x}\overline{\phi_{m}(y;E)}^T
V(x,y)
\phi_p(y;E)e^{i\xi_p(E)x}dydx.
\end{align}
Now we decompose the potential $V$ as
\begin{equation}\label{eq:decom of V}
    V(x,y) = \sum_{k} \sum_{i=0}^3 v_{k,i}(x) \tilde{\varphi}_k(y) \sigma_i,
\end{equation}
where $\{\tilde{\varphi}_k\}$ denotes a basis in $y$-direction and we recall $\sigma_i$ ($i=0,1,2,3$) are the Pauli matrices. Denoting the Fourier transform of $v_{k,i}$ by $\hat{v}_{k,i}(\xi) := \int v_{k,i}(x)e^{-i\xi x} dx$, we have the following relation between the linearized scattering data and the Fourier coefficients of the potential,
\begin{equation} \label{eq:forward data}    
    \hat{S}^{\mathrm{lin}}_{m,p}(E;V) = -\frac{i|E|}
{\sqrt{|\xi_m(E)|\,|\xi_p(E)|}}
    \sum_{k,i} 
    \hat{v}_{k,i}(\xi_{m,p}(E))
    \int 
    \overline{\phi_m(y;E)}^{T} \sigma_i \phi_p(y;E) \tilde{\varphi}_k(y) dy,
\end{equation}
where we denote
\begin{align}\label{ximp}
    \xi_{m,p}(E) := \xi_m(E) - \xi_p(E).
\end{align}
Hereafter, we write $\hat{S}^{(\cdot)}(E)$ for $\hat{S}^{(\cdot)}(E;V)$ whenever no confusion arises. 
Unless otherwise stated, we assume $\xi>0$, since the reality of $v_{k,i}(x)$ implies
$
    \hat{v}_{k,i}(-\xi)=\overline{\hat{v}_{k,i}(\xi)}.
$

\section{Main results}
\label{sec:main}
In this section, we establish uniqueness and stability for scalar
potentials. Theorem~\ref{thm:sclar invertible for single xi} gives
fixed-frequency invertibility of the linearized scattering map, and
Corollary~\ref{cor:stability} provides the corresponding Fourier-domain
stability estimate. We then obtain local nonlinear reconstruction of
finite-dimensional scalar potentials near $V=0$ in
Theorem~\ref{thm:scalar invertible for finite dimension V}. The general
Hermitian problem, subject to the gauge obstruction in
Proposition~\ref{prop:compact-gauge}, is treated in Appendix~\ref{app:full}.

To understand where information on $\hat v_{k,i}(\xi)$ is encoded in \eqref{eq:forward data}, we first state the following result on the range of $\xi_{m,p}(E)$ in~\eqref{ximp}.

\rev{\begin{lemma}\label{lemma:Emp}
Fix $(n,q)\in\N^2\setminus\{(0,0)\}$ and
$\xi\in\mathbb R\setminus\{0,\pm\sqrt{2|n-q|}\}$. There are propagating
indices $m,p$ with $|m|=n$, $|p|=q$, and an energy $E$ such that
$\xi_{m,p}(E)=\xi$. The energy and branches are as follows.
\begin{enumerate}[(i)]
\item If $n,q\geq1$, then
\begin{equation}\label{eq:Emp}
    E=\pm E_{n,q}(\xi),\qquad
    E_{n,q}(\xi):=
    \sqrt{\frac{\xi^2}{4}+(n+q)+\frac{(n-q)^2}{\xi^2}}.
\end{equation}
The branch signs can be chosen as
\begin{equation*}
\begin{cases}
\epsilon_m=-\epsilon_p=\sgn (\xi),
& |\xi|>\sqrt{2|n-q|},\\
\epsilon_m=\epsilon_p=\sgn\bigl (\xi\cdot(q-n)\bigr),
& 0<|\xi|<\sqrt{2|n-q|}.
\end{cases}
\end{equation*}
\item If $n\geq1$ and $q=0$, then $p=(0,-)$ and there is only one energy,
\begin{equation}\label{eq:Emp-zero-right}
    E=\frac{\xi}{2}+\frac{n}{\xi},\qquad
    \epsilon_m=\sgn\!\left(\frac{\xi^2-2n}{\xi}\right).
\end{equation}
\item If $n=0$ and $q\geq1$, then $m=(0,-)$ and there is only one energy,
\begin{equation}\label{eq:Emp-zero-left}
    E=-\frac{\xi}{2}-\frac{q}{\xi},\qquad
    \epsilon_p=-\sgn\!\left(\frac{\xi^2-2q}{\xi}\right).
\end{equation}
\end{enumerate}
In the two exceptional cases, $|E|$ is still given by the right-hand side
of~\eqref{eq:Emp}, but the sign of $E$ is fixed by the ordering of the zero
mode and by $\xi$.
\end{lemma}}

\begin{proof}
When $n,q\geq1$, both longitudinal momenta have the two branches
$\epsilon\sqrt{E^2-2n}$. Solving
$\epsilon_m\sqrt{E^2-2n}-\epsilon_p\sqrt{E^2-2q}=\xi$
gives~\eqref{eq:Emp}; the stated signs follow by comparing $|\xi|$ with
$\sqrt{2|n-q|}$. Since the expression depends only on $E^2$, both signs of
$E$ are admissible.

If $q=0$, equation~\eqref{ximp} becomes
\begin{equation*}
    E+\epsilon_m\sqrt{E^2-2n}=\xi.
\end{equation*}
Squaring after isolating the square root gives
$E=\xi/2+n/\xi$, and the unsquared equation fixes $\epsilon_m$ as in
\eqref{eq:Emp-zero-right}. The case $n=0$ is obtained in the same way from
$-E-\epsilon_p\sqrt{E^2-2q}=\xi$. The excluded values are precisely zero
frequency and the thresholds at which one of the selected modes ceases to be
propagating.
\end{proof}
    The relation given in Lemma~\ref{lemma:Emp} shows how the accessible values of $\xi$ are constrained by the energy $E$ and the indices $m=(n,\epsilon_m),p=(q,\epsilon_p)$. 
This leads to the following observations:
\begin{enumerate}[label=(\roman*)]
    \item For low energy levels $|E|<\sqrt{2}$, the only accessible value is $\xi=0$. 
    Scattering information in this regime is restricted to the $x$-average of $V$
    and yields essentially no useful data about the spatial oscillations of the perturbation $V$.
    \item We deduce from \eqref{eq:Emp},~\eqref{eq:Emp-zero-right} and~\eqref{eq:Emp-zero-left}, the lower bound
    $E^2 \;\ge\; \frac{\xi^2}{4} + (n+q)$,
    which implies that the energy must grow at least linearly with $|\xi|$. 
    Hence, the high-frequency components of $V$, encoding its fine spatial details, are only visible through high-energy scattering data.
    \item  The low frequency components $|\xi|\ll1$ of the potential $V$ may be reached in two possible ways:
    \begin{enumerate}
        \item When $n\neq q$, equation~\eqref{eq:Emp},~\eqref{eq:Emp-zero-right} and~\eqref{eq:Emp-zero-left}, shows that $E\sim |n-q|/|\xi| \to \infty$, so accessing very small $|\xi|$ also requires arbitrarily high energy. 
        \item   When $n=q\geq1$, one can reach small $|\xi|$ at finite energy, but only through reflection data (with $\epsilon_m\neq \epsilon_p$). However, such reflection data is not sufficient to fully recover the potential $V$. For example, when $V$ is odd in the $y$-direction, then, for any $m=(n,\epsilon_m), p=(q,\epsilon_p)$ such that $n=q$,
        \begin{equation}
            \int \bar{\phi}_{m}(y;E)V(x,y)\phi_p(y;E)dy=0,
        \end{equation}
         Thus, by~\eqref{eq:forward data}, $\hat{S}_{m,p}=0$, i.e., these terms of the forward data contain no information about $V$.   
         
 Therefore, for low-frequency $\xi$, complete recovery of $V$ requires high-energy $E\sim|\xi|^{-1}$ scattering data, which is quite different from the setting of scattering data for $H=H_0+V$ with $H_0$ a constant-coefficient operator.
    \end{enumerate}
\end{enumerate}
{For the topologically trivial scalar waveguide associated with \(-\Delta+y^2\), treated in Appendix \ref{app:linearized-harmonic-waveguide}, the lowest-index transverse modes propagate in both directions, and the reflection coefficients allow us to probe
nonzero longitudinal Fourier frequencies as the energy varies. This is the main difference with the case of  the 
Dirac waveguide, where this lowest-mode
back-scattering channel $(0,+)$ is absent.}

\paragraph{Reconstruction of a scalar potential.}

Using the frequency relations in Lemma~\ref{lemma:Emp}, we now study scalar
potentials of the form
    \begin{align}\label{scalar v decomposition}
        V(x,y)=\sum_{k=0}^N v_k(x)\tilde{\varphi}_k(y)\sigma_0,
    \end{align}
with $\tilde{\varphi}_{k}(y)=2^{\frac{1}{2}}\pi^{\frac{1}{4}}\varphi_k(\sqrt{2}y)$ a scaled Hermite function.
%From \eqref{eq:forward data}, the scattering coefficient $\hat{S}$ is then related to $\hat{v}$ by 
{We need to introduce} the following $3$-tensor
\begin{align*}
   \langle \varphi \rangle_{(i,j;k)} = \int \varphi_i(y) \varphi_j(y) \tilde{\varphi}_k(y) dy,\quad \forall\,i,j,k \in \N
\end{align*}
where ${\varphi_i}$ (${\tilde{\varphi}_k}$) is the (rescaled) Hermite function. %introduced in~\eqref{eq:Hermite functions}, and  represents the $y$-direction basis function used in the decomposition of $V$ in~\eqref{eq:decom of V}.
In the decomposition~\eqref{scalar v decomposition} of $V$, $n$ {is} either finite or infinite. Denote the Fourier transform of $v_k$ as $\hat{v}_k(\xi)$. {To recover $\hat{v}_k$, we consider the following (partial) scattering information for each $\xi>0$ and $s\in \N$,
}
\begin{equation*}
    \widetilde S_s^{(\cdot)}(\xi)=
    \begin{cases}
        \displaystyle
        -\frac{\xi}{\sqrt2}\,
        \widehat S_{1+,1-}^{(\cdot)}(E_0(\xi)),
        &s=0,\\[1.2ex]
        \displaystyle
        -\sqrt{\frac{\xi^2}{2s}-1}\,
        \widehat S_{s+,0-}^{(\cdot)}(E_s(\xi)),
        &0<s<\frac{\xi^2}{2},\\[1.2ex]
        \displaystyle
        -\sqrt{1-\frac{\xi^2}{2s}}\,
        \widehat S_{s-,0-}^{(\cdot)}(E_s(\xi)),
        &s>\frac{\xi^2}{2}.
    \end{cases}
\end{equation*} 
\begin{align*}
        E_0(\xi)=\sqrt{\frac{\xi^2}{4}+2},\quad 
        E_s(\xi)=\sqrt{\frac{\xi^2}{4}+s+\frac{s^2}{\xi^2}}, \quad s\in\N^+.
\end{align*}
This subset of scattering data is sufficient to uniquely reconstruct the potential while remaining amenable to explicit inversion formulas. 
We will see from the numerical illustrations in Section~\ref{sec:incomplete data} that other scattering data may be used and, in some cases, improve reconstructions.

To measure the size of the Hermite coefficients of $V$ and of the
corresponding scattering data $S$, we introduce the following weighted $\ell^1$ norms.
For $N\in\N^+\cup\{\infty\}$ and a sequence $a=(a_s)_{s=0}^N$, we set
\begin{equation}\label{eq:weight l_1 norm }
    \|a\|_{\mathcal{V}_N}
    =\sum_{s=0}^N\frac{|a_s|}{\sqrt{s!}},
    \qquad
    \|a\|_{\mathcal{S}_N}
    =\sum_{s=0}^N\frac{2^{s/2}}{\sqrt{s!}}\,
    |a_s|.
\end{equation}
For finite $N$, we regard $\mathcal V_N$ and $\mathcal S_N$
as $\mathbb C^{N+1}$ equipped with the norms
in~\eqref{eq:weight l_1 norm }.
For $N=\infty$, they denote the corresponding weighted
$\ell^1$ spaces. In this case, invertibility refers to the
bounded extension of the coefficient map to these spaces.
The weighted norms in \eqref{eq:weight l_1 norm } are chosen so that the
linearized reconstruction is not only injective but also stable. More precisely,
for each admissible longitudinal Fourier frequency \(\xi\), the linearized map is
bounded from \(\mathcal V_N\) to \(\mathcal S_N\) and admits a bounded inverse.
The following theorem gives a fixed-frequency well-posedness result for the
linearized inverse problem. The corresponding global Fourier-domain stability
estimate is obtained in Corollary~\ref{cor:stability} by integrating this estimate with respect to \(\xi\).

\begin{theorem}\label{thm:sclar invertible for single xi}
    The linearized scattering map for all $\xi \in \R^{+}$ except for the countable set $\{\sqrt{2k} : k \in \N^+\}$
 and $N\in\N^+\cup \{\infty\}$, \begin{align*}
        \mathcal{L}^{\mathrm{lin}}(\xi):  \mathcal V_N\longrightarrow\mathcal S_N
    \quad \big(\hat{v}_k(\xi)\big)_{k=0}^N
    \mapsto 
    \big(\tilde{S}^{\mathrm{lin}}_k(\xi)\big)_{k=0}^N,
    \end{align*}
    is invertible and there exist constants $C_1$ and $C_2$ independent of $\xi$ and $N$, such that,
    \begin{align}\label{eq:y norm equa for T V}
    C_1 \|\hat{v}(\xi)\|_{\mathcal{V}_N}\le \|\tilde{S}^{\mathrm{lin}}(\xi)\|_{\mathcal{S}_N}\le C_2\|\hat{v}(\xi)\|_{\mathcal{V}_N}.
    \end{align}
    \end{theorem}
\begin{proof}
Fix $\xi \in \R^+ \setminus \{\sqrt{2K}: K \in \N^+\}$. 
By Lemma~\ref{lemma:Emp} and \eqref{eq:forward data}, for all $s\in\N^+$ we have
\begin{eqnarray}
    -i\tilde{S}^{\mathrm{lin}}_s(\xi)&=&\sum _{k=0}^N\langle\varphi\rangle_{(0,s;k)}\hat{v}_k(\xi),
    \label{eq:47}\\
     -i\tilde{S}^{\mathrm{lin}}_{0}(\xi)&=&\sum_{k= 0}^N(\langle\varphi\rangle_{(0,0;k)}+\langle\varphi\rangle_{(1,1;k)})\hat{v}_k(\xi).\label{eq:160}
\end{eqnarray}
By direct computation,
\begin{equation}\label{eq:145}
   \langle\varphi\rangle_{(0,s;k)}=
    \begin{cases}
        (-1)^{\frac{s-k}{2}}2^{\frac{k}{2}-s}\sqrt{\dfrac{s!}{k!}}\dfrac{1}{(\frac{s-k}{2})!}, & s-k\in 2\N,\\[1ex]
        0, & \text{otherwise}.
    \end{cases}
\end{equation}
Set
\[
    b_s:=-i\widetilde S_s^{\mathrm{lin}}(\xi),
    \qquad
    \mu_s:=\sum_{k=0}^N
    \langle\varphi\rangle_{(0,s;k)}\widehat v_k(\xi),
    \qquad 0\leq s\leq N.
\]
By~\eqref{eq:47}, $b_s=\mu_s$ for $s\geq1$.
Moreover, the Hermite recurrence gives
\[
    \varphi_1^2=\varphi_0^2+\sqrt2\,\varphi_0\varphi_2.
\]
Hence, for $N\geq2$, equation~\eqref{eq:160} implies
\[
    b_0=2\mu_0+\sqrt2\,\mu_2,
    \qquad
    \mu_0=\frac{b_0-\sqrt2\,b_2}{2}.
\]
For $N=1$, the corresponding relations are
$b_0=\frac32\mu_0$ and $b_1=\mu_1$.
In either case, the transformation between $b$ and $\mu$
is invertible and satisfies
\[
    \frac23\|\mu\|_{\mathcal S_N}
    \leq \|b\|_{\mathcal S_N}
    \leq 2\|\mu\|_{\mathcal S_N}.
\]

We next analyze the map from $\widehat v$ to $\mu$.
Define
\[
    a_s:=\frac{\widehat v_s(\xi)}{\sqrt{s!}},
    \qquad
    z_s:=\frac{2^{s/2}\mu_s}{\sqrt{s!}}.
\]
Formula~\eqref{eq:145} gives
\[
    z_s
    =\sum_{\ell=0}^{\lfloor s/2\rfloor}
    \frac{(-1/2)^\ell}{\ell!}\,a_{s-2\ell}.
\]
Let $R_N$ denote the shift by two indices,
\[
    (R_Na)_s=
    \begin{cases}
        a_{s-2},&s\geq2,\\
        0,&s=0,1.
    \end{cases}
\]
Then $\|R_N\|_{\ell^1\to\ell^1}\leq1$ and
\[
    z=e^{-R_N/2}a,
    \qquad
    a=e^{R_N/2}z.
\]
These identities hold in finite dimensions and on $\ell^1$
when $N=\infty$, since both operator exponential series
converge in operator norm. Consequently,
\[
    e^{-1/2}\|\widehat v(\xi)\|_{\mathcal V_N}
    \leq \|\mu\|_{\mathcal S_N}
    \leq e^{1/2}\|\widehat v(\xi)\|_{\mathcal V_N}.
\]
Combining the two estimates and using
$\|b\|_{\mathcal S_N}
  =\|\widetilde S^{\mathrm{lin}}(\xi)\|_{\mathcal S_N}$,
we obtain~\eqref{eq:y norm equa for T V}, for example with
\[
    C_1=\frac23e^{-1/2},
    \qquad
    C_2=2e^{1/2}.
\]
The explicit inverses above also prove invertibility.
\end{proof}
{Integrating the fixed-frequency stability estimate~\eqref{eq:y norm equa for T V} with respect to the Fourier variable $\xi$, we obtain the following global Fourier-domain stability estimate.}
% To obtain the stability in the complete Fourier domain in $\R_y$, we integrate~\eqref{eq:y norm equa for T V} with respect to $\xi$ in Theorem~\ref{thm:sclar invertible for single xi}.
For simplicity, we denote, $\forall\,m = (n, \epsilon_m)$,
\begin{align*}
    \Lambda_n(E) &= \sqrt{E^2 - 2n}, \quad
    \Xi_m(E) = E + \epsilon_m \Lambda_n(E). \quad
    % \Xi_{m,p}(E) = \sqrt{\Xi_m(E) \Xi_p(E)}.
\end{align*}
\begin{corollary}\label{cor:stability}
    By integrating~\eqref{eq:y norm equa for T V} with respect to $\xi$ in Theorem~\ref{thm:sclar invertible for single xi}, we obtain the following stability estimate for some positive constants $C_1$ and $C_2$:
    \begin{equation}
         \begin{aligned}
C_1 \sum_{s=0}^\infty\int_{0}^\infty \frac{|\hat{v}_s(\xi)|}{\sqrt{s!}}\,d\xi
\le& 
2\sqrt{2}\int_{\sqrt{2}}^\infty
E\big|\widehat S_{1+,1-}^{\mathrm{lin}}(E)\big|\,dE
\\&+
\sum_{s=1}^{\infty}\frac{2^{s/2}}{\sqrt{s!}}
\int_{\sqrt{2s}}^\infty
\sqrt{\frac{2}{\Lambda_s(E)}}
\left[
\frac{\Xi_{s-}(E)}{\sqrt{\Xi_{s+}(E)}}
\big|\widehat S_{s-,0-}^{\mathrm{lin}}(E)\big|
+
\frac{\Xi_{s+}(E)}{\sqrt{\Xi_{s-}(E)}}
\big|\widehat S_{s+,0-}^{\mathrm{lin}}(E)\big|
\right]dE\\
\le& C_2 \sum_{s=0}^\infty \int_{0}^\infty \frac{|\hat{v}_s(\xi)|}{\sqrt{s!}}\,d\xi.
\end{aligned}
    \end{equation}
\end{corollary}
\begin{proof}
    Take $N=\infty$ in Theorem~\ref{thm:sclar invertible for single xi}, substitute the following change of variable into~\eqref{eq:y norm equa for T V},
    \begin{equation}
        \begin{cases}
        \displaystyle
            \xi =2\Lambda_1(E_0),\qquad \ \  \left\lvert\frac{d\xi}{dE_0}\right \rvert=\frac{2E_0}{\Lambda_1(E_0)},  &s=0,\\
             \displaystyle
            \xi =E_s+\Lambda_s(E_s), \quad \left\lvert\frac{d\xi}{d E_s}\right \rvert=\frac{\Xi_{s+}(E_s)}{\Lambda_s(E_s)}, &0<s<\frac{\xi^2}{2},\\
            \displaystyle
            \xi =E_s-\Lambda_s(E_s), \quad \left\lvert\frac{d\xi}{d E_s}\right \rvert=\frac{\Xi_{s-}(E_s)}{\Lambda_s(E_s)}, &s>\frac{\xi^2}{2},         
        \end{cases}
    \end{equation}
    and integrate \eqref{eq:y norm equa for T V} to obtain the result.
\end{proof}

{\begin{remark}
The above analysis also applies to the topologically trivial operator
\(-\Delta+y^2\). Indeed, after replacing the Dirac linearized scattering
coefficients by the scalar waveguide coefficients, the remaining
well-posedness and stability proof is direct: it only uses the
Hermite expansion of \(V\) and the invertibility of the associated Hermite
triple tensor. We record the corresponding Green function, linearized
scattering formula, and stability statement in
Appendix~\ref{app:linearized-harmonic-waveguide}.
\end{remark}}

\paragraph{Nonlinear reconstruction of finite-dimensional scalar potentials.}

Assume that the scalar potential belongs to the finite-dimensional real
vector space
\begin{equation}\label{eq:scalar potential decom}
    V(x,y)
    =
    \sum_{k=0}^N\sum_{j=0}^r
    v_{k,j}e_j(x)\widetilde{\varphi}_k(y)\sigma_0,
    \qquad
    v=(v_{k,j})\in\mathbb R^d,
\end{equation}
where
\(
    d=(N+1)(r+1)
\)
and the functions $e_j$ are real-valued.

Let $\xi_0,\ldots,\xi_r$ be distinct admissible frequencies, and assume
that the Fourier evaluation matrix
\begin{equation}\label{eq:Fourier evaluation matrix}
    A=\big(\widehat e_j(\xi_\ell)\big)_{\ell,j=0}^r
\end{equation}
is invertible over $\mathbb C$. Define the finite-dimensional scattering
data map
\begin{equation}\label{eq:finite-dimensional-nonlinear-map}
    \mathcal S:\mathbb R^d\longrightarrow\mathbb C^d,
    \qquad
    \mathcal S(v)
    =
    \big(\widetilde S_k(\xi_\ell;V)\big)_{
        \substack{0\leq k\leq N\\0\leq\ell\leq r}}.
\end{equation}
\begin{theorem}\label{thm:scalar invertible for finite dimension V}
Assume that the nonlinear scattering map $\mathcal S$ in
\eqref{eq:finite-dimensional-nonlinear-map} is continuously differentiable
in a neighborhood of $v=0$. Suppose that the matrix $A$ in
\eqref{eq:Fourier evaluation matrix} is invertible and that, for every
$\ell=0,\ldots,r$, the fixed-frequency linearized scattering map
\[
    \mathcal L^{\mathrm{lin}}(\xi_\ell):
    \mathbb C^{N+1}\longrightarrow\mathbb C^{N+1}
\]
is invertible.

Then $\mathcal S$ is locally injective at $v=0$. More precisely, there
exist a neighborhood $U\subset\mathbb R^d$ of $0$ and a real-linear map
\(
    P:\mathbb C^d\longrightarrow\mathbb R^d
\)
such that
\[
    P\circ\mathcal S:U\longrightarrow P(\mathcal S(U))
\]
is a $C^1$ diffeomorphism onto an open subset of $\mathbb R^d$.
Consequently, $\mathcal S|_U$ has a $C^1$ inverse on its image.
After possibly shrinking $U$, there also exists $C>0$ such that
\[
    |v-v'|
    \leq
    C\|\mathcal S(v)-\mathcal S(v')\|_{\mathbb C^d},
    \qquad v,v'\in U.
\]
\end{theorem}
\begin{proof}
For $v=(v_{k,j})\in\mathbb R^d$, define
\[
   \mathcal A:\mathbb R^d\longrightarrow\mathbb C^d,\qquad (\mathcal A v)_{k,\ell}
    :=
    \widehat v_k(\xi_\ell)
    =
    \sum_{j=0}^r v_{k,j}\widehat e_j(\xi_\ell).
\]
Since the Fourier evaluation matrix $A$ is invertible over $\mathbb C$,
the linear map \(\mathcal A\)
is injective. 

The derivative of the nonlinear scattering map at $v=0$ factors as
\[
    D\mathcal S(0)
    =
    \left(
        \bigoplus_{\ell=0}^r
        \mathcal L^{\mathrm{lin}}(\xi_\ell)
    \right)
    \circ\mathcal A.
\]
Because $\mathcal A$ is injective and each
$\mathcal L^{\mathrm{lin}}(\xi_\ell)$ is invertible over $\mathbb C$,
the linear map
\[
    D\mathcal S(0):\mathbb R^d\longrightarrow\mathbb C^d
\]
is injective. Equivalently, its real representation
\[
    J_{\mathbb R}:\mathbb R^d\longrightarrow\mathbb R^{2d},
    \qquad
    J_{\mathbb R}v
    =
    \begin{pmatrix}
        \operatorname{Re}D\mathcal S(0)v\\
        \operatorname{Im}D\mathcal S(0)v
    \end{pmatrix},
\]
has rank $d$.
Hence $J_{\mathbb R}$ admits a real-linear left inverse. For example, one
may take
\(
    P_{\mathbb R}
    =
    (J_{\mathbb R}^{\mathsf T}J_{\mathbb R})^{-1}
    J_{\mathbb R}^{\mathsf T}.
\)
Identifying $\mathbb C^d$ with $\mathbb R^{2d}$, let
\(
    P:\mathbb C^d\longrightarrow\mathbb R^d
\)
be the corresponding real-linear map. Then
\[
    P\,D\mathcal S(0)=I_{\mathbb R^d}.
\]

Define
\(
    F(v):=
    P\big(\mathcal S(v)-\mathcal S(0)\big).
\)
It follows that
\[
    DF(0)=P\,D\mathcal S(0)=I_{\mathbb R^d}.
\]
The inverse function theorem therefore shows that $F$ is a local
$C^1$ diffeomorphism near $v=0$.

If $v,v'\in U$ and $\mathcal S(v)=\mathcal S(v')$, then
$F(v)=F(v')$, and hence $v=v'$. Thus $\mathcal S$ is locally injective.
Moreover, on $\mathcal S(U)$ its inverse is given by
\[
    \mathcal S|_U^{-1}(z)
    =
    F^{-1}\big(P(z-\mathcal S(0))\big).
\]
After shrinking $U$ if necessary, the local Lipschitz estimate follows
from the boundedness of $D(F^{-1})$ and of $P$.
\end{proof}

A concrete admissible choice can be constructed from translates of a
single real-valued function. Let
\[
    g\in C_c^\infty((-\delta,\delta);\mathbb R),
    \qquad
    \int_{\mathbb R}g(x)\,dx\neq0,
\]
and choose $x_*\in\mathbb R$ and $h>0$ such that
\[
    [x_*+jh-\delta,x_*+jh+\delta]
    \subset (x_L,x_R),
    \qquad j=0,\ldots,r.
\]
Define
\begin{equation}\label{eq:example basis and freq}
    e_j(x):=g(x-x_*-jh),
    \qquad j=0,\ldots,r.
\end{equation}
These functions are real-valued and compactly supported in
$(x_L,x_R)$.

Since $\widehat g(0)\neq0$, there exists $\delta_0>0$ such that
$\widehat g(\xi)\neq0$ whenever $|\xi|<\delta_0$. Choose
$a,\eta>0$ such that
\[
    0<a<a+r\eta<\min\{\delta_0,\sqrt{2}\},
    \qquad
    rh\eta<2\pi,
\]
and set
\[
    \xi_\ell:=a+\ell\eta,
    \qquad \ell=0,\ldots,r.
\]
In particular, all the frequencies $\xi_\ell$ are nonzero and avoid the
exceptional set $\{\sqrt{2n}:n\in\mathbb N^+\}$.

With the Fourier transform convention
\(
    \widehat f(\xi)=\int_{\mathbb R}e^{-i\xi x}f(x)\,dx,
\)
we have
\[
    \widehat e_j(\xi_\ell)
    =
    e^{-i\xi_\ell(x_*+jh)}\widehat g(\xi_\ell).
\]
Consequently, the Fourier evaluation matrix
\(
    A=\big(\widehat e_j(\xi_\ell)\big)_{\ell,j=0}^r
\)
can be written as
\[
    A
    =
    \operatorname{diag}\!\left(
        e^{-i\xi_\ell x_*}\widehat g(\xi_\ell)
    \right)_{\ell=0}^r
    \big(z_\ell^j\big)_{\ell,j=0}^r,
    \qquad
    z_\ell:=e^{-ih\xi_\ell}.
\]
The second factor is a Vandermonde matrix. The condition
$rh\eta<2\pi$ implies that the numbers
$z_0,\ldots,z_r$ are pairwise distinct. Therefore,
\[
    \det A
    =
    \left(
        \prod_{\ell=0}^r
        e^{-i\xi_\ell x_*}\widehat g(\xi_\ell)
    \right)
    \prod_{0\leq p<q\leq r}(z_q-z_p)
    \neq0.
\]
Thus the Fourier evaluation map is injective on the real coefficient
space, and the conclusion of
Theorem~\ref{thm:scalar invertible for finite dimension V} applies.

\section{Adjoint method and TR matrices} \label{sec:adjoint}
 In this section, we present an adjoint-based optimization approach to reconstruct the potential $V$, which is compactly supported in $[x_L,x_R]\times\Rm$, using the observed TR matrix $T^{ob}$. We restrict our observation of the TR matrix to the index set 
$M^{ob} \subset M \times M$ and choose the weighted Frobenius norm for the optimization objective $\Pi^T(V)$. {The formulation is written for the finite-interval TR coefficients, and in the numerical experiments we restrict the observation set to the propagating coefficients corresponding to the far-field scattering matrix \(S\).}
Let $M^{ob} \subset M \times M$ denote the index set of observed scattering pairs $(m,p)$. Then
\begin{equation}\label{eq:optimization objection for adp}
     \Pi^T =\sum_{(n-,p) \in M^{ob}} w_{n-,p}|\alpha^p_{n-}(x_L) - T^{\mathrm{ob}}_{n-,p}|^2+\sum_{(n+,p) \in M^{ob}}
                    w_{n+,p}|\alpha^p_{n+}(x_R) - T^{\mathrm{ob}}_{n+,p}|^2,
\end{equation}
   
where
$\alpha^p_m$ are defined in~\eqref{eq:alpha ^p _m} and $w_{m,p}>0$ are the weights. We denote 
\[
M_{\mathrm{in}}^{\mathrm{ob}}
:=
\left\{
p\in M:\text{ there exists }m\text{ such that }(m,p)\in M^{\mathrm{ob}}
\right\}.
\] For each $p\in M_{\mathrm{in}}^{\mathrm{ob}}$, we further denote
\begin{align*}
M_p:=\{m\in M:(m,p)\in M^{ob}\},
\qquad
\Pi_p^T:=\sum_{m\in M_p} w_{m,p}\,|\alpha_m^p-T^{\mathrm{ob}}_{m,p}|^2.
\end{align*} In the adjoint formulation, the selection of observation modes $M^{ob}$ is arbitrary. Therefore, the proposed algorithm works naturally when restricting the observation to the scattering matrix $S^{ob}$ or the partial scattering data used in Theorem~\ref{thm:sclar invertible for single xi}. 

\subsection{Adjoint-based optimization}\label{sec:Adjoint-based optimization}
As a discretization, we assume $V$ follows a generalization of~\eqref{eq:decom of V}:
\begin{equation}
    \label{eq:function expansion of perturbation V}
    V(x,y)=\sum_{A}\kappa_AV_{A}(x,y).
\end{equation}
%which is a .
We first differentiate \eqref{eqn:psi_dcomp} with respect to $\kappa_A$ and obtain,
\begin{equation}\label{eq:27}
    (H_0-E+V)\frac{\partial \psi^p_{out}}{\partial \kappa_A}+V_A(\psi^p_{in}+\psi^p_{out})=0.
\end{equation}
We then differentiate $\Pi^T_p$ in \eqref{eq:optimization objection for adp} with respect to $\kappa_A$. The first term satisfies
\begin{equation}\label{eq:differential first term}
\begin{aligned}
    &\frac{\partial}{\partial \kappa_A}\sum_{n-\in M_p}w_{n-,p}\big|\alpha^p_{n-}(x_L)-T^{ob}_{n-,p}\big|^2
    =\sum_{n-\in M_p}2w_{n-,p}\Re\Big(\frac{\partial\overline{\alpha^p_{n-}(x_L)}}{\partial \kappa_A}\big(\alpha^p_{n-}(x_L)-T^{ob}_{n-,p}\big)\Big)\\
    &=\sum_{n-\in M_p}2w_{n-,p}\Re\Big(\Big\langle\frac{\partial\psi^p_{out}(x_L,\cdot)}{\partial \kappa_A},
    \frac{e^{i\xi_{n-}x_L}}{1-|P_n|^2}\big(\phi_{n-}(\cdot)-P_n\phi_{n+}(\cdot)\big)\Big\rangle_y
    \big(\alpha^p_{n-}(x_L)-T^{ob}_{n-,p}\big)\Big)\\
    &=2\Re\Big\langle\frac{\partial\psi^p_{out}}{\partial \kappa_A},\,f_{-}^p\Big\rangle_{(x,y)},
\end{aligned}
\end{equation}
and similarly the second term satisfies
\begin{equation}\label{eq:differential second term}
    \frac{\partial}{\partial \kappa_A}\sum_{n+\in M_p}w_{n+,p}\big|\alpha^p_{n+}(x_R)-T^{ob}_{n+,p}\big|^2
    =2\Re\Big\langle\frac{\partial\psi^p_{out}}{\partial \kappa_A},\,f_{+}^p\Big\rangle_{(x,y)}.
\end{equation}

Here, $f_{-}^p$ and $f_{+}^p$ are given by:
% (supported on the boundaries $x=x_L$ and $x=x_R$, respectively):
\begin{align*}
    f_{-}^p(x,y)
    &=\delta(x-x_L)\sum_{n-\in M_p} w_{n-,p}\big(\alpha^p_{n-}(x_L)-T^{ob}_{n-,p}\big)
    \frac{\phi_{n-}(y)-P_n\phi_{n+}(y)}{1-|P_n|^2},\label{eq:f_minus}\\
    f_{+}^p(x,y)
    &=\delta(x-x_R)\sum_{n+\in M_p} w_{n+,p}\big(\alpha^p_{n+}(x_R)-T^{ob}_{n+,p}\big)
    \frac{\phi_{n+}(y)-\overline{P_n}\phi_{n-}(y)}{1-|P_n|^2}. %\label{eq:f^plus}
\end{align*}

Let $f^p=f_{-}^p+f_{+}^p$. 
We seek $g^p$ as a distributional solution of the adjoint problem
\begin{equation}\label{eq:adjoint problem dist}
    (H_0-E+V)^* g^p = f^p,
\end{equation}
where by self-adjointness, in fact $(H_0-E+V)^* = H_0-E+V$. For any outgoing wave $h$ such that $(H_0-E)h=0$ outside $[x_L,x_R]$, the following identity holds:
\begin{equation}\label{eq:adjoint green identity}
    \langle g^p,(H_0-E+V)h\rangle_{(x,y)}=\langle (H_0-E+V)^* g^p,h\rangle_{(x,y)}=\langle f^p,h\rangle_{(x,y)}.
\end{equation}
Applying \eqref{eq:adjoint green identity} with $h=\partial_{\kappa_A}\psi^p_{out}$ and using~\eqref{eq:27},
we obtain
\begin{equation}
    \label{eq:adjoint outcome}
    \begin{aligned}
    \frac{\partial\Pi^T_p}{\partial \kappa_A}
    &=2\Re\langle f^p,\partial_{\kappa_A}\psi^p_{out}\rangle_{(x,y)}
    =2\Re\langle (H_0-E+V)^*g^p,\partial_{\kappa_A}\psi^p_{out}\rangle_{(x,y)}\\
    &=2\Re\langle g^p,(H_0-E+V)\partial_{\kappa_A}\psi^p_{out}\rangle_{(x,y)}
    =-2\Re\langle g^p,V_A\psi^p\rangle_{(x,y)}.
\end{aligned}
\end{equation}

\subsection{Integral formulation for adjoint problem}\label{sec:Integral formulation}
 Equation~\eqref{eq:adjoint outcome} provides the derivative of the objective functional
with respect to the expansion coefficients. To solve~\eqref{eq:adjoint problem dist}, we decompose $g^p$ by two parts $g^p=g^p_{in}+g^p_{out}$, where $g^p_{in}$ and $g^p_{out}$ satisfy,
 \begin{align}
     &(H_0-E)g^p_{in}=f^p,\label{eq:g^p_{in}}\\
     &(H_0-E+V)g^p_{out}=-Vg^p_{in}\label{eq:g^p_{out}}.
 \end{align}
Equation~\eqref{eq:g^p_{in}} is solved explicitly below and
once an explicit expression for $g^p_{in}$ is obtained, 
the numerical methods proposed in~\cite{bal2023asymmetric} 
allow us to compute $g^p_{out}$.

\paragraph{Computation of $g^p_{in}$.} To {obtain an} explicit expression for $g^p_{in}$, we introduce the incoming Green's function $G_{in}(x,y;x_0,y_0)$ solution of:
\begin{equation}
\label{eq:def of incoming Green's func}
    (H_0-E)G_{in}=\delta(x-x_0)\delta(y-y_0)I.
\end{equation}
As in \eqref{eq:G_out + -}, $G_{in}$ admits the following explicit form
\begin{align}
    G_{in}=\begin{pmatrix}
        (D_x+E)G_{in,+} & \fa G_{in,-}\\
        \fa^*G_{in,+} & (-D_x+E)G_{in,-}
    \end{pmatrix},
\end{align}
%with
\begin{align*}
    &G_{in,-}(x,y;x_0,y_0;E)=\sum_{n= 0}^{\infty}\frac{1}{2\theta_n(E)}e^{-\theta_n(E)|x-x_0|}\varphi_n(y)\varphi_n(y_0),\\
    &G_{in,+}(x,y;x_0,y_0;E)=\sum_{n= 1}^{\infty}\frac{1}{2\theta_{n}(E)}e^{-\theta_{n}(E)|x-x_0|}\varphi_{n-1}(y)\varphi_{n-1}(y_0).
\end{align*}

Then following \eqref{eq:g^p_{in}}, $g^p_{in}$ is given by
\begin{align}
    g^p_{in}(x,y)=\int G_{in}(x,y;x_0,y_0;E)f^p(x_0,y_0)dx_0dy_0.
\end{align}
By direct computation, we obtain the following explicit expression for $g^p_{in}$ on the interval $(x_L,x_R)$,
\begin{equation}\label{eq:g^p decomp}
    g^p_{in}(x,y)=\sum_{m\in M_p}\alpha_{m}[g^p_{in}](x)\phi_{m}(y),   
\end{equation}
%where
\begin{align}
        \alpha_{n-}[g^p_{in}](x)&=w_{n-,p}(\alpha^p_{n-}(x_L)-T^{ob}_{n-,p})\frac{\frac{E}{\theta_n}-P_n\frac{c_{n+}}{c_{n-}}(\frac{E}{\theta_n}+i)}{1-|P_n|^2}e^{i\xi_{n-}(x-x_L)},    \label{eq:coeff of g^p in 1}
\\
        \alpha_{n+}[g^p_{in}](x)&=w_{n+,p}(\alpha^p_{n+}(x_R)-T^{ob}_{n+,p})\frac{\frac{E}{\theta_n}-\overline{P_n}\frac{c_{n-}}{c_{n+}}(\frac{E}{\theta_n}-i)}{1-|P_n|^2}e^{i\xi_{n+}(x-x_R)}.   \label{eq:coeff of g^p in 2}
      \end{align}

Note that $g^p_{in}$ is an incoming wave function on the interval $(x_L,x_R)$ and that~\eqref{eq:g^p_{out}} is of the same form as the outgoing wave equation~\eqref{eq:outgoing solution}. Thus, given the incoming conditions $\alpha_{+}[g^p_{in}](x_L)$ and $\alpha_{-}[g^p_{in}](x_R)$, we may use the numerical methods proposed in~\cite{bal2023asymmetric} 
to compute $g_{p,\mathrm{out}}$ and $g^p$ on the interval $[x_L, x_R]$. The corresponding numerical procedures are outlined in Alg.~\ref{alg:single slab} and Alg.~\ref{alg:eigenfunctions in a slab} in the appendix.

\subsection{The algorithm for adjoint-based optimization}

We now summarize the adjoint-based iterative algorithm to solve the inverse scattering problem.
Given an initial guess of the potential coefficients $\{\kappa_A^{(0)}\}$, the method proceeds by
alternating between a forward scattering solve, an adjoint solve, and a gradient-based update
of the coefficients.

At the $\iter$-th iteration, the current approximation of the potential is
$
V^{(\iter)}(x,y) := \sum_A \kappa_A^{(\iter)} V_A(x,y).
$ The corresponding TR matrix,
denoted by $ T^{(\iter)}$, is computed by solving the forward scattering problem.
The mismatch between $ T^{(\iter)}$ and the reference data $T^{\mathrm{ref}}$ defines the
objective functional $\Pi^T( V^{(\iter)})$.

To efficiently compute the gradient of $\Pi^T$ with respect to the coefficients $\kappa_A$,
we solve the adjoint problem derived in Section~\ref{sec:Adjoint-based optimization}, with the explicit gradient
formula given by~\eqref{eq:adjoint outcome}. The numerical solution of the adjoint problem is carried out using the integral formulation described in
Section~\ref{sec:Integral formulation}, and implemented via
Algorithms~\ref{alg:single slab} and~\ref{alg:eigenfunctions in a slab} in the appendix. The resulting gradient is then used to update the coefficients $\kappa_A$ by a gradient descent
step.  The complete procedure is summarized in Algorithm~\ref{alg_itera}.
\begin{algorithm}[H]
	\caption{Adjoint method in inverse scattering problem}  \label{alg_itera}
	\begin{algorithmic}[1]
	 \Require Initial guess for coefficients $\kappa_A$ of the potential; Interval $I=[x_L,x_R]$ where the potential is supported;
    % Level of binary merging $L$; 
    Reference TR matrix $T^{\mathrm{ref}}$; Observation configuration $M^{ob}$; Number of iterations $\iter_{\max}$, Update size $\eta$.
    \Ensure Approximated potential $V\approx \sum_A \kappa_A V_A$.
	\For{$\iter$ in $0\rightarrow \iter_{\max}$}
            \State Compute TR matrix $T^{(\iter)}$ for potential $V^{(\iter)}=\sum_A\kappa_A^{(\iter)}V_A$ restricted to the interval $[x_L,x_R]$ by Algorithm \ref{alg:eigenfunctions in a slab}.
            \For{$p \in M_{\mathrm{in}}^{\mathrm{ob}} $}
            \State Use Algorithm \ref{alg:eigenfunctions in a slab} to recover $\psi^p$ on interval $I$.
            \For{$m \in M_p$}
            \State Compute the coefficient $\alpha_{n-}[g^p_{in}](x_R)$ and $\alpha_{n+}[g^p_{in}](x_L)$ using equation \eqref{eq:coeff of g^p in 1} and \eqref{eq:coeff of g^p in 2}.
            \EndFor
            \State Use the potential field ${V}^{(\iter)}$ and  coefficients $\alpha_{n-}[g^p_{in}](x_R)$,  $\alpha_{n+}[g^p_{in}](x_{L})$ as input for Algorithm \ref{alg:eigenfunctions in a slab} to recover $g^p$ on interval $I$.
            \For{each $A$}
        \State Compute and store
        $G_{A,p}:=\dfrac{\partial\Pi_p^T}{\partial\kappa_A}$
        using~\eqref{eq:adjoint outcome}.
        \EndFor
            \EndFor
           \For{each A}
                \State Calculate $ \frac{\partial \Pi^T}{\partial \kappa_A}=\sum_{p\in M_1}\frac{\partial \Pi^T_p}{\partial \kappa_A}$.
                \State Update $\kappa_A$ by $\kappa_A^{(\iter+1)}=\kappa_A^{(\iter)}-\eta\frac{\partial \Pi^T}{\partial \kappa_A}$.
            \EndFor
            \EndFor
	\end{algorithmic}
\end{algorithm}

\section{Numerical results of the adjoint method}\label{sec:num}
This section presents numerical experiments using the adjoint-based inverse scattering method. In Section~\ref{sec:numer validation}, we validate
Algorithm~\ref{alg_itera} for the reconstruction of compactly supported potentials, and examine its robustness with respect to measurement noise {and energy level}. In Section~\ref{sec:incomplete data}, we investigate the reconstruction performance
when only partial scattering data are available, and in particular those described  in Theorem~\ref{thm:sclar invertible for single xi}. 
Section~\ref{sec:nonrecoverable} first illustrates the gauge obstruction for compactly supported general Hermitian potentials: although the components $v_2$ and $v_3$ are not separately determined, their gauge-invariant combination $\partial_xv_2+\partial_yv_3$ can be reconstructed. We then consider additional non-reconstructibility phenomena for restricted potential classes with non-compact transverse support.

In the following numerical experiments, we denote $(n_x, n_y)$ as the discretization configuration and $n_E$ as the number of observed energy levels. {Unless mentioned otherwise, the observed data consist of the far-field scattering coefficients for propagating modes \(m,p\in M(E)\) with transverse indices \(n,q\le n_y\), i.e. \[M^{0}(E)=\{(m,p)\in M(E)\times M(E)\mid m=(n,\epsilon),\,p=(q,\delta),\ n,q \le n_y\},\]} the weights in the optimization
objective \eqref{eq:optimization objection for adp} are
$w_{m,p}\equiv 1$, and when applying Algorithm~\ref{alg_itera}, we initialize the potential as ${V}^{(0)}=0$.
{These choices also play a regularizing role in the numerical inverse problem. 
The reconstruction is performed in a finite-dimensional approximation space, 
and only finitely many modes and energy levels are used in the objective 
functional. Thus the discretization parameters $(n_x,n_y)$, the observation 
set $M^{\rm ob}$, and the energy window determine the effective degrees of 
freedom and the amount of scattering information used in the inversion.} 
%In this sense, the numerical reconstruction may be viewed as involving a low-pass filtering or truncation of the recoverable components.
To quantify the reconstruction performance at $\iter$-th  step, we
define the normalized data misfit
{\begin{align}\label{eq:data_misfit}
    \mathcal{S}(\iter)
    :=\frac{\displaystyle\sum_{i=1}^{n_E}\sum_{(m,p)\in M^{0}(E^{i})}
    w_{m,p}\,|{S}^{(\iter)}_{m,p}(E^i)-S_{m,p}^{\mathrm{ref}}(E^i)|^2}
            {\displaystyle\sum_{i=1}^{n_E}\sum_{(m,p)\in M^{0}(E^i)}
            w_{m,p}\,|{S}^{(0)}_{m,p}(E^i)-S_{m,p}^{\mathrm{ref}}(E^i)|^2},
\end{align}
where ${S}^{(\iter)}$ denotes the corresponding scattering matrix.} 
\subsection{Convergence and  stability}\label{sec:numer validation}
In this subsection, we set $n_x=10$, $n_y=15$, $n_E=18$, and energy levels
$\{E^i\}_{i=1}^{n_E}$ uniformly distributed in $[1.5,15]$. The potential is assumed
to be compactly supported on $[x_L,x_R]=[-0.4,0.4]$.

The reference potential is discretized using a tensor-product Legendre--Hermite basis,
\begin{equation}\label{eq:V_ref decom L-H}
    V^{\mathrm{ref}}(x,y)
    =\sum_{j=0}^{n_x}\sum_{k=0}^{n_y}
    v_{j,k}^{\mathrm{ref}}\, P_{j}(x)\,\varphi_k(y)\,\sigma_0,
\end{equation}
where $P_{j}(x)$ denotes the $j$-th Legendre polynomial on the interval
$
[x_L,x_R],
$
and $\varphi_k(y)$ is the $k$-th Hermite function. {Here the same cutoff
$n_y$ is used both for the Hermite expansion of the potential and
for the observed transverse mode indices in the scattering data. This choice
is consistent with the reconstruction mechanism in Theorem~\ref{thm:sclar invertible for single xi}, where
the potential coefficients \((\hat v_k(\xi))_{k=0}^N\) are recovered from
scattering data whose transverse mode indices are retained up to the same
cutoff \(N\).} During the optimization procedure, the potential at the $\iter$-th iteration is represented in the same basis as
\(
    {V}^{(\iter)}(x,y)
    =\sum_{j,k} v_{j,k}^{(\iter)}\,P_{j}(x)\,\varphi_k(y)\,\sigma_0.
\) Since all mode interactions of the form
$
\int_{\mathbb R} \varphi_m^{T}(y) V(x,y)\varphi_p(y)\,dy
$
are evaluated using Gauss--Hermite quadrature rules in the numerical implementation,
the potential $V(x,y)$ effectively enters the forward and adjoint solvers through its
values at the Gauss--Hermite quadrature points. A Hermite parameterization therefore yields a discretization consistent with the numerical forward model. We measure the relative error of the reconstructed potential at $\iter$-th  step by
\begin{align}\label{def:err}
    \mathcal{E}(\iter)
    :=\frac{\displaystyle\sum_{j,k}\frac{1}{2j+1}
    \big(v_{j,k}^{(\iter)}-v_{j,k}^{\mathrm{ref}}\big)^2}
            {\displaystyle\sum_{j,k}\frac{1}{2j+1}
            \big(v_{j,k}^{\mathrm{ref}}\big)^2}.
\end{align}

\refstepcounter{experiment}
\label{exp:1}
\textbf{Experiment \theexperiment.} As our first example, we reconstruct the scalar potential
\begin{align*}
    V = V_0(x,y)\,\sigma_0,
\end{align*}
where the function $V_0$ is obtained by interpolating an image of letter 'H' onto the Legendre--Hermite basis in~\eqref{eq:V_ref decom L-H}. To validate the stability of Algorithm~\ref{alg_itera}, we reconstruct the potential from the {noisy scattering matrices}, which are set by,
{\begin{align*}
\tilde{S}^{\text{ref}}_{m,p}=S_{m,p}^{(0)}+(1+\sigma z_{m,p})(S^{\text{ref}}_{m,p}-S_{m,p}^{(0)}),\quad (m,p)\in M^{ob},
\end{align*}}
where $z_{m,p}\sim\mathcal{N}(0,1)$ are i.i.d., $\sigma$ is the varying level of noise. The same measurement noise realization is used with various multiplicative factors $\sigma$. {We terminate the adjoint iteration at $\iter_{\max}=600$, since by this stage the reconstruction saturates at a noise-dependent level. 
This termination also prevents the reconstruction from continuing to fit noise-dominated components of the measured scattering data.}

\begin{figure}[H]
    \centering
    \includegraphics[width=0.45\linewidth]{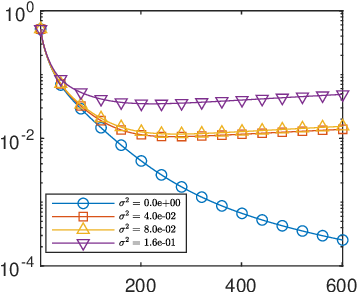}
    \caption{
    Relative reconstruction errors $\mathcal{E}$ against iterations
    for different noise levels $\sigma$.
    }
    \label{fig:exp1_noiseH_error}
\end{figure}

Figure~\ref{fig:exp1_noiseH_error} depicts the evolution of the relative reconstruction error 
$\mathcal{E}(\iter)$
as a function of the number of iterations $\iter$ for various noise levels. In the noise-free case, the relative reconstruction error decreases steadily
over the iterations.
In the presence of noise, the error $\mathcal{E}$ saturates after an initial decay. The saturation level increases with the noise variance $\sigma^2$, indicating a noise-dependent error floor.  
We also monitor the normalized scattering data misfit $\mathcal{S}(\iter)$ defined in
\eqref{eq:data_misfit}.
At the final iteration $\iter_{\max}=600$, the data misfit decreases to the order
of $10^{-5}$ in the noise-free case.
In the presence of noise, $\mathcal{S}(\iter)$ exhibits a clear saturation
behavior: it stabilizes at the order of $10^{-4}$ for $\sigma^2=4\times10^{-2}$,
around $10^{-3}$ for $\sigma^2=8\times10^{-2}$, and at the order of $10^{-2}$ for
$\sigma^2=1.6\times10^{-1}$.
This noise-dependent misfit floor is consistent with the plateau observed in the
reconstruction error $\mathcal{E}$ and reflects the limitation imposed by
measurement noise.

\begin{figure}[H]
    \centering
    \begin{subfigure}[t]{0.31\textwidth}
        \centering
        \includegraphics[width=\linewidth]{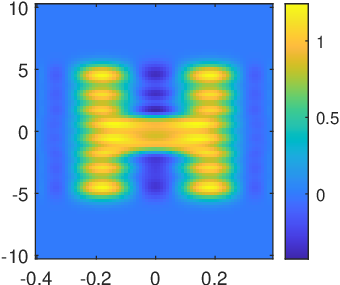}
        \caption{Reference}
    \end{subfigure}
    \hfill
    \begin{subfigure}[t]{0.31\textwidth}
        \centering
        \includegraphics[width=\linewidth]{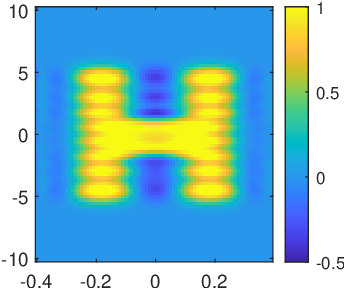}
        \caption{Reconstruction}
    \end{subfigure}
    \hfill
    \begin{subfigure}[t]{0.31\textwidth}
        \centering
        \includegraphics[width=\linewidth]{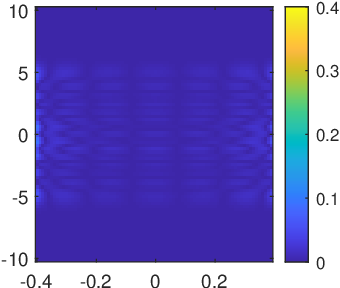}
        \caption{Absolute error}
    \end{subfigure}
    \caption{
    Reconstruction of the potential from scattering data with no noise
    at iteration $600$ .
    }
    \label{fig:exp1_noiseH_reconstruction}
\end{figure}

\begin{figure}[H]
    \centering
    % --- row 1: reconstructions ---
    \begin{subfigure}[t]{0.31\textwidth}
        \centering
        \includegraphics[width=\linewidth]{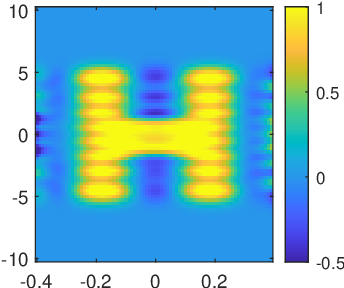}
        \caption{$\sigma^2 = 4\times 10^{-2}$}
    \end{subfigure}
    \hfill
    \begin{subfigure}[t]{0.31\textwidth}
        \centering
        \includegraphics[width=\linewidth]{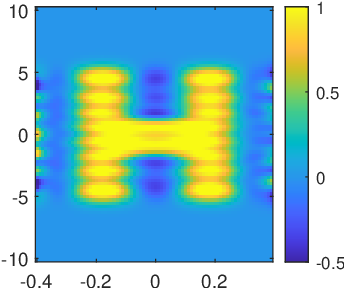}
        \caption{$\sigma^2 = 8\times 10^{-2}$}
    \end{subfigure}
    \hfill
    \begin{subfigure}[t]{0.31\textwidth}
        \centering
        \includegraphics[width=\linewidth]{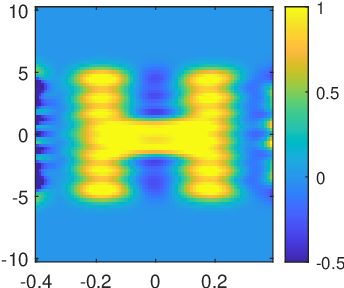} 
        \caption{$\sigma^2 = 1.6\times 10^{-1}$}
    \end{subfigure}

    \vspace{2mm}

    % --- row 2: absolute errors ---
    \begin{subfigure}[t]{0.31\textwidth}
        \centering
        \includegraphics[width=\linewidth]{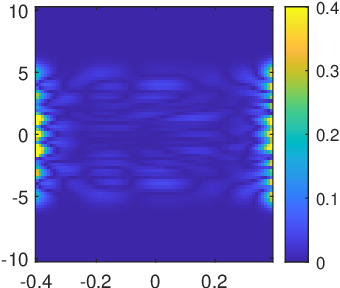}
        \caption{Abs. error}
    \end{subfigure}
    \hfill
    \begin{subfigure}[t]{0.31\textwidth}
        \centering
        \includegraphics[width=\linewidth]{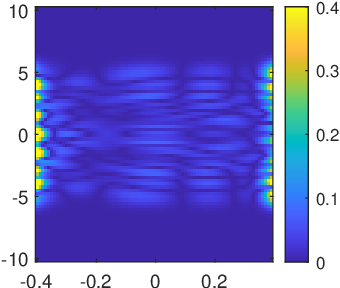}
        \caption{Abs. error}
    \end{subfigure}
    \hfill
    \begin{subfigure}[t]{0.31\textwidth}
        \centering
        \includegraphics[width=\linewidth]{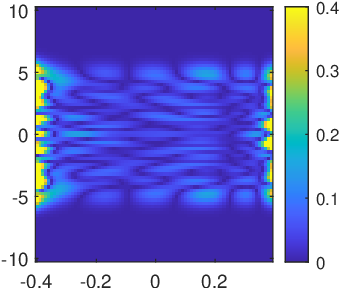} 
        \caption{Abs. error}
    \end{subfigure}

    \caption{
    Reconstructions (top) and absolute errors (bottom)
    under different noise levels, shown at iteration $\iter_{\max}=600$.
    }
    \label{fig:exp1_noiseH_multilevel}
\end{figure}

Figure~\ref{fig:exp1_noiseH_reconstruction} presents the reconstruction at final iteration in the absence of noise. The recovered potential closely matches the reference,
and the absolute error remains small across the domain. Figure~\ref{fig:exp1_noiseH_multilevel} compares the reconstructions and absolute errors for different noise levels. As the noise level increases, the absolute error gradually increases and exhibits more pronounced artifacts.

\begin{figure}[H]
        \centering
        \includegraphics[width=0.45\linewidth]{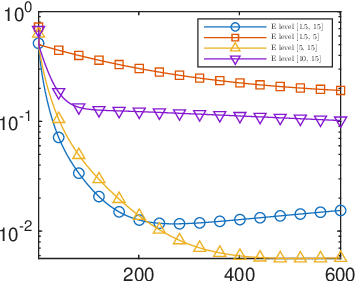}
        \caption{Relative reconstruction errors \(\mathcal E\) against iterations for the noisy setting \(\sigma^2=8.0\times 10^{-2}\) and different energy intervals.}
     \label{fig:energy_sensitivity}
\end{figure}

{\paragraph{Sensitivity to the energy interval.}

We also examine the sensitivity of the reconstruction to the choice of the energy interval in the noisy setting. We fix the noise level at \(\sigma^2=8.0\times 10^{-2}\) and keep the same discretization parameters as above, while changing only the observed energy interval. In addition to the baseline interval \([1.5,15]\), we consider the intervals \([1.5,5]\), \([5,15]\), and \([10,15]\).

Figure~\ref{fig:energy_sensitivity} reports the corresponding relative reconstruction errors. The baseline interval \([1.5,15]\) and the interval \([5,15]\) lead to comparable reconstruction errors, with \([5,15]\) giving a slightly smaller final error in this example. This indicates that the reconstruction is relatively stable when the lowest-energy part of the data is removed. By contrast, the lower-energy interval \([1.5,5]\) gives a noticeably worse reconstruction, with the error remaining much larger throughout the iteration. This is consistent with the frequency-accessibility discussion in Section~\ref{sec:main}: a restricted low-energy window provides less information about the longitudinal Fourier components of the potential. The narrower high-energy interval \([10,15]\) also gives a less accurate reconstruction, suggesting that both the location and the width of the energy interval affect the amount of usable scattering information.}

\subsection{Incomplete scattering data}\label{sec:incomplete data}
In this section, we numerically validate the discussion presented before Theorem~\ref{thm:sclar invertible for single xi} on the choice of scattering data to recover the Fourier coefficients $\hat v_k(\xi)$ of the potential at a fixed frequency $\xi$.
In the theoretical analysis, the injectivity and stability results are
established using a specific subset of scattering coefficients associated with
pairs $(m,p)$ for which the mode indices satisfy
$\{n,q\}=\{0,s\}$.
This particular selection corresponds to the observation set
{
\begin{equation*}
    M^A=\{(m,p)\in M^0\mid m=(n,\epsilon),\ p=(q,\delta),\ \{n,q\}=\{0,s\} \},
\end{equation*}}
that leads to explicit and tractable inversion formulas as in Theorem~\ref{thm:sclar invertible for single xi}. On the other hand, the magnitude of the energy in
Lemma~\ref{lemma:Emp}
\begin{align*}
    |E_{n,q}(\xi)|=\sqrt{\frac{\xi^2}{4}+(n+q)+\frac{(n-q)^2}{\xi^2}},
\end{align*}
indicates that, for fixed $\xi$ and fixed $n+q$, the required energy magnitude
increases as $|n-q|$ increases. When one index is zero, the ordering of the
channels fixes the sign of $E$ as in~\eqref{eq:Emp-zero-right}--\eqref{eq:Emp-zero-left}.
Due to the finite range of $E$ in the numerical simulation, scattering coefficients associated with indices
$(n,q)$ that are closer to each other may exhibit improved conditioning. This motivates us to consider a larger collection of scattering
coefficients 
{\begin{equation*}
    M^B=\{(m,p)\in M^0\mid m=(n,\epsilon),\ p=(q,\delta),\ |n-q|<1\},
\end{equation*}}
and to compare their reconstruction performance with the theoretically motivated choice of $M^A$.

\refstepcounter{experiment}
\label{exp:2}
\textbf{Experiment \theexperiment.}
To speed up convergence when using incomplete scattering data, we
choose a smaller discretization with $n_x=6$, $n_y=10$, and a shorter interval
$[x_L,x_R]=[-0.2,0.2]$.
All other numerical configurations and the relative error $\mathcal{E}$ of the reconstructed potential are the same as in~\textbf{Experiment~\ref{exp:1}}.
We consider the scalar reference potential
\begin{equation*}
    V^{\mathrm{ref}}(x,y)
    =\sum_{j=0}^{n_x}\sum_{k=0}^{n_y}
    v_{j,k}^{\mathrm{ref}}\, P_{j}(x)\,\varphi_k(y)\,\sigma_0,
\end{equation*}
where the coefficients are interpolated from the function
$f(x,y)=\pi^{\frac{1}{4}}\cos\left(\frac{2\pi x}{x_R-x_L}\right)e^{-\frac{y^2}{2}}$.

\begin{figure}[H]
        \centering
        \includegraphics[width=0.45\linewidth]{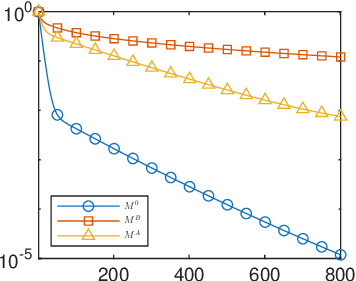}
        \caption{  {Relative reconstruction errors $\mathcal{E}$ against iterations}}
    
    \label{fig:eg11}
\end{figure}

Figure~\ref{fig:eg11} reports the relative reconstruction error $\mathcal{E}$
as a function of the iteration number for different choices of scattering data.
For the observation set $M^0$, where all coefficients of the {scattering data} are used, the error decays most rapidly, as expected.
For the observation set associated with $M^A$, the reconstruction remains stable with
$\mathcal{E}$ decreasing steadily to values below $10^{-2}$ with about $500$ iterations. This confirms that the data subset employed in the theoretical analysis is not only sufficient for uniqueness but also effective for practical reconstruction. With the observation set $M^B$, which consists of scattering coefficients not used in the theoretical stability analysis, the error decreases more slowly and remains larger throughout the iterations.
Nevertheless, the reconstruction  exhibits a consistent decay of $\mathcal{E}$, indicating that such coefficients still contribute to numerical robustness. 

The behavior of the normalized scattering data misfit $\mathcal{S}$ (residual error) is similar to the relative reconstruction errors $\mathcal{E}$. Using the full observation set $M^0$, the data misfit decreases to about $10^{-5}$ at final iteration. For the observation set $M^A$, the misfit shows a monotone decay to about $10^{-2}$, while for the observation set $M^B$, the data misfit decreases more slowly to about  $10^{-1}$.

\subsection{Non-reconstructible potential}\label{sec:nonrecoverable}
In this subsection, we aim to numerically validate several intrinsic
non-reconstructibility phenomena predicted by the scattering theory.
In particular, we consider classes of potentials for which the scattering
matrix does not contain sufficient information to uniquely determine the
potential, even in the absence of noise.

We begin with the gauge obstruction for a compactly supported general
Hermitian potential. Theorem~\ref{thm:well-posed-ness full} shows that the
linearized scattering data determine $v_0$ and $v_1$ separately, whereas the
last two components are determined only through
$\partial_xv_2+\partial_yv_3$. We use the nonlinear adjoint reconstruction to
illustrate the same gauge-invariant structure numerically.

\refstepcounter{experiment}
\label{exp:3}
\textbf{Experiment \theexperiment.}
We now consider a Hermitian potential with three nonzero Pauli components
to illustrate the gauge obstruction described in
Proposition~\ref{prop:compact-gauge}. The reference potential is
\begin{equation}\label{eq:general-Hermitian-reference-H}
    V^{\mathrm{ref}}(x,y)
    =V_0^{\mathrm{ref}}(x,y)\sigma_0
     +V_2^{\mathrm{ref}}(x,y)\sigma_2
     +V_3^{\mathrm{ref}}(x,y)\sigma_3,
    \qquad V_1^{\mathrm{ref}}=0.
\end{equation}
As in \textbf{Experiment~\ref{exp:1}}, each of the three nonzero components
$V_0^{\mathrm{ref}}$, $V_2^{\mathrm{ref}}$, and $V_3^{\mathrm{ref}}$ is
obtained by interpolating the same image of the letter `H' onto the
Legendre--Hermite basis.
We keep $v_1=0$ throughout the reconstruction and optimize only the
$\sigma_0$, $\sigma_2$, and $\sigma_3$ components.
The discretization, energy levels, observation set, optimization weights,
and initialization are the same as in \textbf{Experiment~\ref{exp:1}} and no
measurement noise is added.

Theorem~\ref{thm:well-posed-ness full} shows that the full linearized
scattering data determine $v_0$ separately, whereas $v_2$ and $v_3$ are
determined only through the gauge-invariant combination
$\partial_xv_2+\partial_yv_3$.
To examine this distinction numerically, we compare the reconstruction
errors for
\[
    f\in\left\{
        v_0,\ \partial_xv_2,\ \partial_yv_3,\
        \partial_xv_2+\partial_yv_3
    \right\}.
\]
Using the same squared-norm convention as in~\eqref{def:err}, we define
the frequency-aggregated relative squared error by
\begin{equation}\label{eq:gauge-frequency-error}
    \mathcal E_f(\iter)
    :=
    \frac{
        \displaystyle\sum_{\xi\in\Xi}
        \big\|\widehat f^{(\iter)}(\xi,\cdot)
        -\widehat f^{\mathrm{ref}}(\xi,\cdot)\big\|_{L^2(\mathbb R_y)}^2
    }{
        \displaystyle\sum_{\xi\in\Xi}
        \big\|\widehat f^{\mathrm{ref}}(\xi,\cdot)\big\|_{L^2(\mathbb R_y)}^2
    }.
\end{equation}
Here $\Xi$ is obtained from a uniform grid with spacing $0.05$ in
$\{\xi:0.25\leq|\xi|\leq6.25\}$ by excluding points near the exceptional
frequencies $\pm\sqrt{2j}$.
Frequencies at which the reference coefficient norm is numerically
negligible are omitted from both sums for the corresponding quantity.
In the Legendre--Hermite representation, the longitudinal Fourier transform
is evaluated using the Legendre coefficients, while transverse
differentiation is performed using the Hermite differentiation matrix.
In particular,
\[
    \widehat{\partial_xv_2}(\xi,\cdot)
    =i\xi\widehat v_2(\xi,\cdot),
    \qquad
    \widehat{\partial_yv_3}(\xi,\cdot)
    =\partial_y\widehat v_3(\xi,\cdot).
\]
Their sum gives the Fourier coefficients of the gauge-invariant quantity
$\partial_xv_2+\partial_yv_3$.

\begin{figure}[H]
    \centering
    \includegraphics[width=0.52\linewidth]{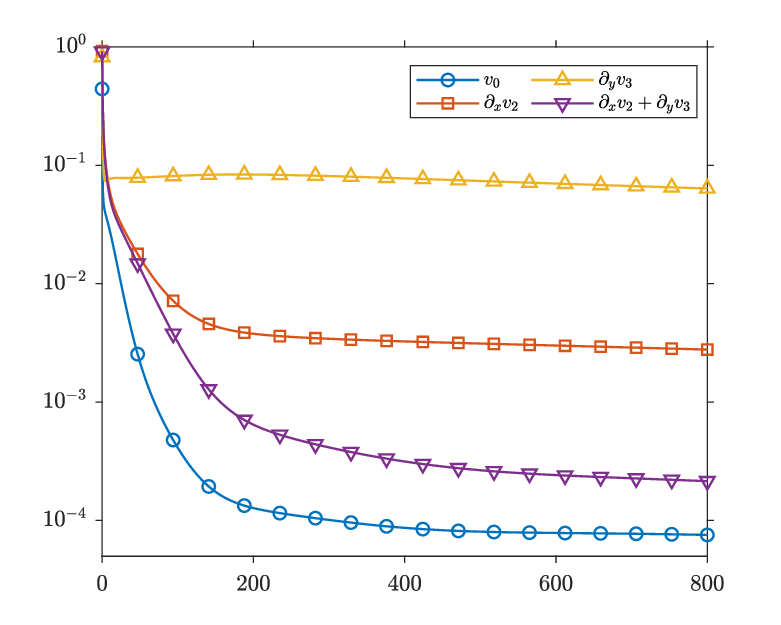}
    \caption{Frequency-aggregated relative squared errors
    \eqref{eq:gauge-frequency-error} for $v_0$, $\partial_xv_2$,
    $\partial_yv_3$, and the gauge-invariant combination
    $\partial_xv_2+\partial_yv_3$ against the recorded optimization steps
    in the noise-free setting.}
    \label{fig:general-Hermitian-gauge-error}
\end{figure}

Figure~\ref{fig:general-Hermitian-gauge-error} displays the resulting
errors over the recorded optimization steps.
The error in $v_0$ decreases rapidly and reaches approximately
$7\times10^{-5}$.
The error in the gauge-invariant combination
$\partial_xv_2+\partial_yv_3$ also decreases, reaching approximately
$2\times10^{-4}$ at the final recorded step.
By contrast, the errors in the two separate terms
$\partial_xv_2$ and $\partial_yv_3$ decrease much more slowly after the
initial transient and remain approximately
$2.5\times10^{-3}$ and $6\times10^{-2}$, respectively.
Thus the relative squared error in the gauge-invariant sum is more than
one order of magnitude smaller than that in $\partial_xv_2$ and more
than two orders of magnitude smaller than that in $\partial_yv_3$.

Although $v_2$ and $v_3$ are not individually identifiable,
their gauge-invariant combination is reconstructed
substantially more accurately in this experiment.
This behavior is consistent with
Proposition~\ref{prop:compact-gauge} and
Theorem~\ref{thm:well-posed-ness full}.

We next consider a case with non-compact support.
\begin{proposition}\label{prop:sigma3_case}
Let $V(x,y) = V_1(x) \sigma_3$. 
Then the scattering matrix for $V$ is diagonal, with transmission entries for each propagating mode given by
\begin{align*}
\hat{S}_{m,m} = e^{-i W(+\infty)}, \qquad W(x) = \int_{-\infty}^x V_1(x')\, dx',
\end{align*}
while all other entries of the {scattering matrix vanish.} In particular, the only information about $V$ that can be recovered from the {scattering matrix} is the total integral
\begin{align*}
W(+\infty) = \int_{-\infty}^{+\infty} V_1(x)\, dx.
\end{align*}
\end{proposition}
\begin{proof}
To see this, let
$
\psi'(x,y) = e^{-i W(x)} \psi(x,y),
$
where $\psi$ is an eigenstate of the unperturbed operator $H_0 - E$. A direct calculation shows
\begin{align*}
(H_0 + V - E)\psi' = e^{-i W(x)} (H_0 - E) \psi = 0.
\end{align*}

By the uniqueness of properly normalized scattering states, this implies that $\psi'$ produces the same outgoing states as the free operator up to a phase factor $e^{-i W(+\infty)}$. Therefore, the transmission entries with the same mode are $e^{-iW(+\infty)}$, and the rest of the entries of the {scattering} matrix vanish.
\end{proof}

The above calculation applies to the original nonlinear inverse problem. The same obstruction persists in the linearized regime. 
 
\refstepcounter{experiment}
\label{exp:4}
\textbf{Experiment \theexperiment.}
To validate Proposition~\ref{prop:sigma3_case}, we consider the reference potential 
\begin{equation*}\label{eqn:nonconsturctible V sigma3}
    V(x,y)=V_1\sigma_3=(x+0.1)\sigma_3.
\end{equation*}
The numerical configurations are set to be the same as in \textbf{Experiment~\ref{exp:2}}.
In the numerical implementation, we discretize $V$ as 
\begin{align*}    V(x,y)
=\sum_{j=0}^{n_x}
    v_{j}^{\text{ref}}\, P_{j}(x)\sigma_3,\end{align*}
where $P_{j}(x)$ is the $j$-th Legendre polynomial on the compact interval 
$[x_L,x_R]=[-0.2,0.2]$.

During the optimization procedure, the potential at the $\iter$-th iteration is represented as
\begin{equation*}
    {V}^{(\iter)}(x,y)
    =\sum_{j} v_{j}^{(\iter)}\,P_{j}(x)\,\sigma_3.
\end{equation*}
We measure the $L_2$ relative error $\mathcal{E}$ of the same form as \eqref{def:err}, while projected to $\sigma_3$,
and we measure the relative error of the average value of the $\sigma_3$ channel by
\begin{align*}
    \mathcal{E}_{\mathrm{avg}}(\iter)=\left(\frac{v^{(\iter)}_{3,0}-v^{\text{ref}}_{3,0}}{v^{\text{ref}}_{3,0}}\right )^2.
\end{align*}

\begin{figure}[H]
    \centering
    \begin{subfigure}[b]{0.45\textwidth}
    \centering
         \includegraphics[width=\linewidth]{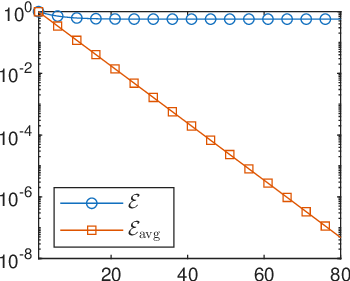}
        \caption{In logarithmic scale.}
        \label{fig:eg9_errVmean_and_errV_all_modes_log} 
    \end{subfigure}
    \hfill
    \begin{subfigure}[b]{0.45\textwidth}
    \centering
         \includegraphics[width=\linewidth]{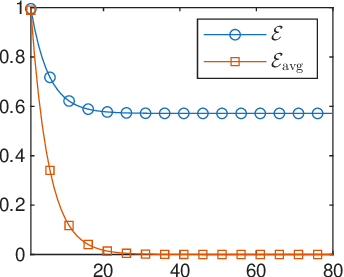}
        \caption{In linear scale.}
        \label{fig:eg9_errVmean_and_errV_all_modes_linear} 
    \end{subfigure}
          \caption{{Reconstruction error measured by $\mathcal{E}$ and $\mathcal{E}_{\mathrm{avg}}$ with respect to iteration steps.}}
              \label{fig:eg9}
\end{figure}

In Figure \ref{fig:eg9}, we present the relative errors $\mathcal{E}$ and $\mathcal{E}_{\mathrm{avg}}$ of the recovered potential. We can see the $L_2$ relative error $\mathcal{E}$ will stay around $0.6$ while the average value can be recovered accurately. This is consistent with Proposition~\ref{prop:sigma3_case}. The normalized scattering data misfit $\mathcal{S}$ 
reaches the order of $10^{-7}$ at the final iteration.

\medskip
We now consider the roles of transmission and reflection scattering data on the reconstruction of a scalar potential, also with unbounded support.

\begin{proposition}\label{nonconstructible}
Let $V(x,y) = V_0(x) I_2$ be a scalar potential. 
In the linearized scattering regime, the only nontrivial entries of the
scattering matrix that depend on $V_0(x)$ are
\begin{enumerate}[label=(\roman*)]
    \item Transmission for propagating modes, {\begin{align*}M^{T}:=\{(m,m)\in M^0\}.\end{align*}}
    \item Reflection between the two propagating directions of modes,  {\begin{align*}M^{R}:=\{(m,p)\in M^0|m=(n,\epsilon), p=(q,\delta), n=q,\epsilon \neq \delta \}.\end{align*}}
\end{enumerate}

All other entries vanish identically for all energies and therefore contain no information about $V_0(x)$.

From the transmission data, we obtain the zero Fourier mode
$
\hat V_0(0) = \int V_0(x)\,dx,
$
while the reflection data determine 
$\hat V_0(\xi)$ for every $\xi \neq 0$ as the energy $E$ varies.  
\end{proposition}
\begin{proof}

To see this, recall that the linearized scattering data satisfy
\begin{equation}\label{forward_data2}
    \hat{S}^{\mathrm{lin}}_{m,p}(E)
    \propto 
    \hat{V}_0\bigl(\xi_{m,p}(E)\bigr)
    \int \overline{\phi_m(y;E)}^{T} \phi_p(y;E)\, dy .
\end{equation}

Let $m = (n,\epsilon_m)$ and $p = (q,\epsilon_p)$.  We consider three cases separately.
\\
(i) If $n \neq q$, then~\eqref{forward_data2} vanishes identically.
\\
(ii)  If $m = p$, then
    \begin{align*}
    \int \overline{\phi_m(y;E)}^{T} \phi_p(y;E)\, dy = 1, 
    \qquad
    \xi_m(E) - \xi_p(E) = 0,
    \end{align*}
so as $E$ varies, the data $\hat{S}^{\mathrm{lin}}_{m,m}$ probe only $\hat{V}_0(0)$, i.e., the mean value of $V_0(x)$ over the interval.
\\
(iii) If $n = q$ but $\epsilon_p \neq \epsilon_m$ for some $n \ge 1$ and $E > \sqrt{2n}$, then
    \begin{align*}
    \int \overline{\phi_m(y;E)}^{T} \phi_p(y;E)\, dy = \frac{\sqrt{2n}}{E} \neq 0,
    \qquad
    \xi_m(E) - \xi_p(E) = \pm 2 \sqrt{E^2 - 2n}.
    \end{align*}
As $E$ varies, $\xi_m(E) - \xi_p(E)$ sweeps all nonzero real values, so the reflection coefficients $\hat{S}^{\mathrm{lin}}_{n+,n-}$ and $\hat{S}^{\mathrm{lin}}_{n-,n+}$ determine $\hat{V}_0(\xi)$ for all $\xi \neq 0$.
\end{proof}

\refstepcounter{experiment}
\label{exp:5}
\textbf{Experiment \theexperiment.} To validate Proposition \ref{nonconstructible} in the nonlinear setting, we consider the potential
\begin{equation}\label{eqn:nonconsturctible V}
    V(x,y)=(x+0.1)\sigma_0,
\end{equation}
and compare the reconstructions from the observation set $M^0$, $M^T$ and $M^R$.
The discretization is identical to that of \textbf{Experiment~\ref{exp:2}} and we measure the $L_2$ relative error $\mathcal{E}$ defined in \eqref{def:err}. 

\begin{figure}[H]
    \centering
    \begin{subfigure}[b]{0.45\textwidth}
        \centering
        \includegraphics[width=\linewidth]{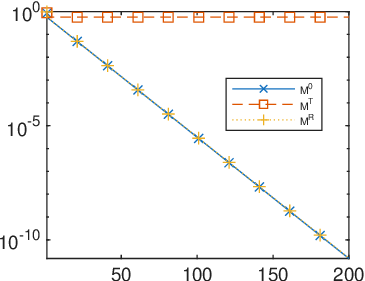}
        \caption{In log scale.}
        \label{fig:Exp4 errV log scale}
    \end{subfigure}
    \hfill
    \begin{subfigure}[b]{0.45\textwidth}
        \centering
        \includegraphics[width=\linewidth]{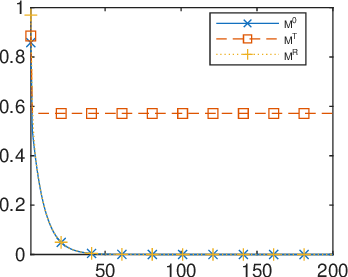}
        \caption{In linear scale.}
        \label{fig:Exp4 errV linear scale}
    \end{subfigure}

    \caption{{Relative errors of recovered potentials by various observations of scattering entries.}}
    \label{fig:eg8-3 errV}
\end{figure}

Figure \ref{fig:eg8-3 errV} depicts the relative errors $\mathcal{E}$ of the reconstruction from various observation sets of scattering coefficients. With full scattering data $M^0$ or only reflection data $M^R$, the relative errors consistently decrease to about $10^{-10}$ after $150$ iterations. However, when recovering the potential using only transmission data $M^T$, the relative error first decreases but then stagnates about $0.6$.
We observe a similar behavior for the scattering data misfit $\mathcal{S}$.
With the full data set $M^0$ and with the reflection-only set $M^R$, the misfit
decreases to about $6\times 10^{-13}$.% by $\iter=200$.

In contrast, using only the transmission data $M^T$, the misfit drops rapidly at first before saturating around $1.5\times 10^{-3}$.

\section{Conclusions} \label{sec:conclu}

This paper presents injectivity and stability results on the inverse scattering theory of Dirac operators with a confining domain wall. \rev{For compactly supported scalar potentials, we prove unique and stable reconstruction in the linearized setting from a subset of the scattering data. For a general Hermitian potential $V=\sum_{j=0}^3v_j\sigma_j$, strict uniqueness fails because of the compactly supported gauge transformation~\eqref{eq:gauge-potential}; the linearized data determine $v_0$, $v_1$, and $\partial_xv_2+\partial_yv_3$, or equivalently the gauge class of $V$. Thus the non-trivial edge topology does not obstruct scalar reconstruction, while the Dirac gauge symmetry gives an unavoidable obstruction for general matrix-valued perturbations.}

We also consider the nonlinear inverse problem, albeit in a restricted setting. We show that scalar potentials represented in a finite-dimensional basis can indeed be uniquely reconstructed under a smallness assumption, as an application of an inverse function theorem. An algorithm based on a standard adjoint method is then presented and used to illustrate and validate our theoretical results by a number of numerical reconstructions based on (synthetic) scattering data.

This inverse scattering problem is an example of a one-dimensional waveguide embedded in a two-dimensional environment, modeling two topological insulators in different phases. An interesting feature of such inverse scattering problems is the necessity to use scattering data for arbitrarily high energies to reconstruct the low-wavenumber structure of the unknown scattering potential. We expect such a feature to persist for a large number of waveguide models, whether or not they are topologically trivial. In particular, it is straightforward to apply the method proposed in this paper to the inverse scattering theory of the Klein Gordon operator $H_{KG}=-\Delta+y^2+V(x,y)$, for $V(x,y)$ compactly supported and real-valued.

\section*{Acknowledgments} 
GB's research is funded in part by NSF grant DMS-230641. ZW's research is funded in part by NTU-SUG and SPMS Collaborative Research Award.

\section*{Data availability statement}
The data that support the findings of this study are available from the corresponding author upon reasonable request.

\appendix

\section*{Appendix}
The appendix provides the following additional details. 
% Section \ref{app:lem}, provides a lemma that allowing us to construct weighted norms on two sequence spaces linked by lower-diagonal linear transforms, such that the induced bijection is continuous and has a continuous inverse.
Section \ref{app:linearized-harmonic-waveguide} concerns the scattering theory of the topologically trivial operator $-\Delta+y^2$.
In Section \ref{app:full}, we identify the gauge kernel of the linearized map for a general Hermitian potential and formulate the corresponding uniqueness result modulo gauge. In Section \ref{app:TRalg}, we provide the algorithm for the computation of transmission reflection matrices in the Dirac model with a linear domain wall.

{\section{Linearized scattering problem for the Klein Gordon operator}
\label{app:linearized-harmonic-waveguide}

This section presents the inverse scattering theory of the operator $-\Delta+y^2$.
For convenience, we use the shifted harmonic oscillator convention
\begin{align}\label{eq:harmonic-waveguide}
-\Delta+y^2-1=-\partial_x^2-\partial_y^2+y^2-1 = -\partial^2_x+\fa^*\fa .
\end{align}
The shift by \(-1\) only translates the spectral parameter and is used to normalize the transverse eigenvalues as \(E_n^2=2n\), matching the threshold notation of the Dirac model.
Let \(\{\varphi_n\}_{n\geq 0}\) be the normalized Hermite functions of
\(L^2(\mathbb R_y)\) as in~\eqref{eq:Hermite functions}.  Then, the transverse operator $h_y=-\partial_y^2+y^2-1$ satisfies, 
\begin{align*}
h_y\varphi_n=2n\varphi_n,
\qquad n=0,1,2,\dots .
\end{align*}

The outgoing Green function is given by,
\begin{align*}
G_0^+(x,y;x',y';E)
=
\sum_{n=0}^{\infty}
\varphi_n(y)\varphi_n(y')
\frac{i}{2\xi_n(E)}
e^{i\xi_n(E)|x-x'|},\quad 
\xi_n(E)
=
\begin{cases}
\sqrt{E^2-2n}, & 2n<E^2,\\[4pt]
i\sqrt{2n-E^2}, & 2n>E^2 .
\end{cases}
\end{align*}
More precisely, separating propagating and evanescent modes gives
\begin{align}\label{eq:explicit form green's function for harmonic waveguide}
\begin{aligned}
G_0^+(x,y;x',y';E)
={}&
\sum_{2n<E^2}
\varphi_n(y)\varphi_n(y')
\frac{i}{2\sqrt{E^2-2n}}
e^{i\sqrt{E^2-2n}|x-x'|}
\\
&+
\sum_{2n>E^2}
\varphi_n(y)\varphi_n(y')
\frac{1}{2\sqrt{2n-E^2}}
e^{-\sqrt{2n-E^2}|x-x'|}.
\end{aligned}
\end{align}
As in the main text, we assume $E^2\notin \{2n:n=0,1,2,\dots\}$.
For a fixed energy $E$,  we define the  propagation index set
\begin{align*}
    P(E):=\{\,m=(n,\epsilon_m) \mid  E^2-2n>0 ,\epsilon_m=\pm 1\},
\end{align*}
and still write $n\pm$ to denote $(n,\pm 1)$.
Consider $-\Delta+y^2-1+V(x,y)$
where \(V\) is compactly supported in the \(x\)
direction. For a fixed incoming mode \(p=q+\in P(E)\), with \(2q<E^2\), 
the unnormalized incoming wave from the left is $u_p^{\mathrm{in}}(x,y;E) = e^{i\xi_qx}\varphi_q(y)$.
The total field \(u_m\) solves
$
(H_0+V-E^2)u_m=0$ and
satisfies $u_p
=
u_p^{\mathrm{in}}
-
R_0(E^2+i0)(Vu_p)$.
Equivalently,
\begin{align*}
u_p(x,y)
=
e^{i\xi_qx}\varphi_q(y)
-
\iint_{\mathbb R^2}
G_0^+(x,y;x',y';E)
V(x',y')u_p(x',y')
\,dx'dy' .
\end{align*}
Under the linearized setting,
\begin{align*}
u_p^\mathrm{lin}(x,y)
=
e^{i\xi_qx}\varphi_q(y)
-
\iint_{\mathbb R^2}
G_0^+(x,y;x',y';E)
V(x',y')u_p^\mathrm{in}(x',y')
\,dx'dy' .
\end{align*}

As \(x\to+\infty\), 
$
e^{i\xi_n|x-x'|}
=
e^{i\xi_nx}e^{-i\xi_nx'}
$
for \(x'\) in the support of \(V\).
Using the explicit Green function, we obtain the asymptotic expansion
\begin{align*}
u_p(x,y)
\sim
e^{i\xi_qx}\varphi_q(y)
-
\sum_{2n<E^2}
\frac{i}{2\xi_n}
e^{i\xi_nx}\varphi_n(y)
\iint_{\mathbb R^2}
e^{-i\xi_nx'}
\varphi_n(y')
V(x',y')u_p(x',y')
\,dx'dy'.
\end{align*}
Thus, for $m=n+$, the normalized transmission coefficient is
\begin{align*}
S_{mp}(E)-\delta_{mp}
=
-
\frac{i}{2\sqrt{\xi_n\xi_ q}}
\iint_{\mathbb R^2}
e^{-i\xi_nx}
\varphi_n(y)
V(x,y)u_p(x,y)
\,dxdy .
\end{align*}
%this is the exact formula for the scattered part \(S(E)-I\).

Under the linearized setting,
the first-order approximation is obtained by replacing \(u_p\) in the integral
formula by
$
u_p^{\mathrm{in}}(x,y;E)
=
e^{i\xi_px}\varphi_p(y).
$
Hence
\begin{align*}
(S(E)-I)^{\mathrm{lin}}_{mp}
=
-
\frac{i}{2\sqrt{\xi_n\xi_q}}
\iint_{\mathbb R^2}
e^{-i(\xi_n-\xi_q)x}
\varphi_n(y)V(x,y)\varphi_q(y)
\,dxdy .
\end{align*}

A similar derivation for reflection coefficients gives the following explicit formulation for linearized scattering data,  $\forall m=(n,\epsilon_m),p=(q,\epsilon_p)\in P(E)$,
\begin{align}\label{eq:linearized scattering for Harmonic waveguide}
    (S(E)-I)^{\mathrm{lin}}_{mp}
=
-
\frac{i}{2\sqrt{\xi_n\xi_q}}
\iint_{\mathbb R^2}
e^{-i\xi_{m,p}(E)x}
\varphi_n(y)V(x,y)\varphi_q(y)
\,dxdy .
\end{align}
where $\xi_{m,p}(E) := \epm(E^2-2n)^{\frac{1}{2}}-\epsilon_p(E^2-2q)^{\frac{1}{2}}$.
Decomposing $V(x,y)=\sum_{k=0}^N
v_k(x)\tilde{\varphi}_k(y)$,
then
\begin{align}
    (S(E)-I)^{\mathrm{lin}}_{m,p}=-
\frac{i}{2\sqrt{\xi_n\xi_q}}\sum_{k=0}^N\hat{v}_k(\xi_{m,p})\int \varphi_n(y)\varphi_q(y)\tilde{\varphi}_k(y) \,dy.
\end{align}
We still take $\tilde{\varphi}_k$ as the scaled Hermite function
$\tilde{\varphi}_k(y)=2^{\frac{1}{2}}\pi^{\frac{1}{4}}\varphi_k(\sqrt{2}y)$
and consider the following (partial) scattering information for each $\xi>0$ and $s\in \N$
\begin{equation*}
   \mathcal{S}_s(\xi) =
    \begin{cases}
        \displaystyle
        \sqrt{\frac{\xi^2}{4}-\frac{s^2}{\xi^2}}\,
        (S(E_s(\xi))-I)^{\mathrm{lin}}_{s+,0-},
        & 0 \le s < \tfrac{\xi^2}{2}, \\[2ex]
        \displaystyle
        \sqrt{\frac{s^2}{\xi^2}-\frac{\xi^2}{4}}\,
         (S(E_s(\xi))-I)^{\mathrm{lin}}_{s-,0-},
        & s > \tfrac{\xi^2}{2},
    \end{cases}\quad E_s(\xi)=\sqrt{\frac{\xi^2}{4}+s+\frac{s^2}{\xi^2}}.
\end{equation*}

The choice of \(E_s(\xi)\) follows the same frequency-matching
principle as in Lemma~\ref{lemma:Emp}. For a fixed target longitudinal Fourier
frequency \(\xi\) and a proper pair of modes \((m,p)\), the energy is
chosen so that
\(
    \xi_{m,p}(E)=\xi .
\)
Thus \(E_s(\xi)\) is determined by the selected scattering entry and
by the Fourier frequency one wishes to recover; it is not an additional
independent parameter. In this way, the scattering coefficient evaluated
at \(E=E_s(\xi)\) probes the Fourier component \(\hat v_k(\xi)\).

The only difference between the present choice of partial data and the
corresponding choice in the Dirac domain-wall model occurs at the lowest
component \(s=0\). In the scalar harmonic waveguide, the lowest
transverse mode has two propagation directions, so the reflection
coefficient \((S(E)-I)^{\rm lin}_{0+,0-}\) can be used to probe
nonzero longitudinal Fourier frequencies. This gives a more direct
choice of the \(s=0\) datum. In the Dirac domain-wall model, the branch
\((0,+)\) is absent, and this lowest-mode backscattering channel is
therefore unavailable; the \(s=0\) component is instead accessed through
the \((1+,1-)\) scattering coefficient.

Still consider the weighted  $l^1$ norm $\|\cdot\|_{\mathcal{V}_N}$ for $\hat{v}(\xi)$ and $\|\cdot\|_{\mathcal{S}_N}$ for $\mathcal{S}(\xi)$ as~\eqref{eq:weight l_1 norm }. With the same derivation in the proof of Theorem~\ref{thm:sclar invertible for single xi}, we obtain that, there exist constants $C_1$ and $C_2$ such that
 \begin{align}\label{eq:y norm equa for S V for Delta+y^2}
    C_1 \|\hat{v}(\xi)\|_{\mathcal{V}_N}\le \|\mathcal{S}(\xi)\|_{\mathcal{S}_N}\le C_2\|\hat{v}(\xi)\|_{\mathcal{V}_N}.
\end{align}
Integrating~\eqref{eq:y norm equa for S V for Delta+y^2} with respect to $\xi$, we obtain the following stability estimate:
\begin{equation*}
         \begin{aligned}
C_1 \sum_{s=0}^\infty\int_{0}^\infty \frac{|\hat{v}_s(\xi)|}{\sqrt{s!}}\,d\xi
\le & 2\int_0^\infty E\,
\left|(S(E)-I)^{\mathrm{lin}}_{0+,0-}\right|\,dE\\&+\sum_{s=1}^\infty \frac{2^{\frac{s}{2}}}{\sqrt{s!}}\int_{\sqrt{2s}}^\infty 
\sqrt{\frac{E}{\sqrt{E^2-2s}}}\Bigg(
(E-\sqrt{E^2-2s})|(S(E)-I)^{\mathrm{lin}}_{s-,0-}|\\&+(E+\sqrt{E^2-2s})|(S(E)-I)^{\mathrm{lin}}_{s+,0-}|
\Bigg)\,dE  \\
\le & C_2 \sum_{s=0}^\infty \int_{0}^\infty \frac{|\hat{v}_s(\xi)|}{\sqrt{s!}}\,d\xi.
\end{aligned}
\end{equation*}
}

\section{General Hermitian potentials: uniqueness modulo gauge}\label{app:full}
Proposition~\ref{prop:compact-gauge} shows that gauge-related potentials
cannot be distinguished by scattering data. We now address the
complementary question of which gauge-invariant quantities the linearized
data determine in a Hermite representation. The coefficient-level argument
below also identifies the kernel at each admissible longitudinal frequency.
Set
\begin{equation}\label{eq:non scalar V}
 V=v_0\sigma _0+v_1\sigma _1+v_2\sigma _2+v_3\sigma _3,
 \qquad
 v_j(x,y)=\sum_{k=0}^Nv_{k,j}(x)\widetilde\varphi_k(y).
\end{equation}

\begin{theorem}\label{thm:well-posed-ness full}
Fix
$
    \xi\in\mathbb R\setminus
    \left(
        \{0\}\cup
        \{\pm\sqrt{2j}:j\in\mathbb N^+\}
    \right).
$
For potentials of the form~\eqref{eq:non scalar V}, the full
linearized scattering data at longitudinal frequency \(\xi\)
determine uniquely
\begin{equation}\label{eq:gauge-invariant-components}
    \widehat v_0(\xi,\cdot),\qquad
    \widehat v_1(\xi,\cdot),\qquad
    \widehat v_2(\xi,\cdot)
    +\frac{1}{i\xi}\partial_y\widehat v_3(\xi,\cdot).
\end{equation}
% In terms of Hermite coefficients, if
% \[
%     \mathbf v_j(\xi)
%     :=
%     \bigl(\widehat v_{k,j}(\xi)\bigr)_{k\geq0},
% \]
% then the data determine uniquely
% \begin{equation}\label{eq:recoverable-Hermite-combination}
%     \mathbf v_0(\xi),\qquad
%     \mathbf v_1(\xi),\qquad
%     \mathbf c(\xi)
%     :=
%     \mathbf v_2(\xi)
%     +\frac{1}{i\xi}D\mathbf v_3(\xi).
% \end{equation}
% Consequently, the ambiguity in the pair
% \((\mathbf v_2,\mathbf v_3)\) is precisely
% \begin{equation}\label{eq:Hermite-gauge-ambiguity}
%     (\mathbf v_2,\mathbf v_3)
%     \longmapsto
%     (\mathbf v_2-D\mathbf h,\,
%      \mathbf v_3+i\xi\mathbf h),
% \end{equation}
% where \(\mathbf h\) is any coefficient sequence for which the
% modified pair belongs to the admissible class.
In the \(x\)-space variables, the determined combination in
\eqref{eq:gauge-invariant-components} is equivalent, for
\(\xi\neq0\), to the determination of
\[
    \reallywidehat{\partial_xv_2+\partial_yv_3}(\xi,\cdot)
    =
    i\xi\widehat v_2(\xi,\cdot)
    +\partial_y\widehat v_3(\xi,\cdot).
\]
This agrees exactly with the gauge invariance stated in
Proposition~\ref{prop:compact-gauge}.
\end{theorem}

\begin{proof}
Fix \(\xi\) as in the statement. We first identify the combination of
\(v_2\) and \(v_3\) that can enter the data. Let \(m,p\) be propagating
modes at an energy \(E\) such that
$
    \xi_m(E)-\xi_p(E)=\xi,
$
and set
\[
    A_j(y):=
    \overline{\phi_m(y;E)}^{T}\sigma_j\phi_p(y;E),
    \qquad j=0,1,2,3.
\]
For \(\ell=m,p\), the transverse profile satisfies
\[
    \bigl(i\sigma_2\partial_y+y\sigma_1
          +\xi_\ell(E)\sigma_3\bigr)\phi_\ell(y;E)
    =E\phi_\ell(y;E).
\]
Using these two equations to differentiate \(A_2\), we obtain
\begin{equation}\label{eq:transverse-gauge-identity}
    \begin{aligned}
    \partial_yA_2
    &={}
    i\overline{\phi_m}^{T}
      \bigl(E-y\sigma_1-\xi_m(E)\sigma_3\bigr)\phi_p
\\
    &\quad
    -i\overline{\phi_m}^{T}
      \bigl(E-y\sigma_1-\xi_p(E)\sigma_3\bigr)\phi_p
\\
    &=-i\bigl(\xi_m(E)-\xi_p(E)\bigr)
      \overline{\phi_m}^{T}\sigma_3\phi_p
     =-i\xi A_3.
\end{aligned}
\end{equation}

It follows from \eqref{eq:forward data} that the contribution of
\(v_2\sigma_2+v_3\sigma_3\) to the Born integral is
\begin{equation}    \label{eq:Born-factorization}
    \begin{aligned}
    \mathcal B^{23}_{m,p}(E)
    &:={}
    \int_{\mathbb R}
    \left[
        \widehat v_2(\xi,y)A_2(y)
        +\widehat v_3(\xi,y)A_3(y)
    \right]dy
\\
    &=
    \int_{\mathbb R}
    \left[
        \widehat v_2(\xi,y)
        +\frac{1}{i\xi}\partial_y\widehat v_3(\xi,y)
    \right]A_2(y)\,dy.
\end{aligned}
\end{equation}

Here we used \eqref{eq:transverse-gauge-identity} and integrated by
parts. 

For fixed \(\xi\), write
\[
    \widehat v_j(\xi,y)
    =\sum_{k\geq0}\widehat v_{k,j}(\xi)
      \widetilde\varphi_k(y),
    \qquad
    \mathbf v_j(\xi)
    :=\bigl(\widehat v_{k,j}(\xi)\bigr)_{k\geq0}.
\]
For a coefficient sequence \(\mathbf a=(a_k)_{k\geq0}\), let \(D\)
denote differentiation in the scaled Hermite basis:
\begin{equation*}
    \partial_y\left(\sum_{k\geq0}a_k\widetilde\varphi_k\right)
    =\sum_{k\geq0}(D\mathbf a)_k\widetilde\varphi_k,
    \qquad
    (D\mathbf a)_k
    =\sqrt{k+1}\,a_{k+1}-\sqrt{k}\,a_{k-1},
\end{equation*}
where \(a_{-1}:=0\); finite sequences are extended by zero. Define
\begin{equation*}
    g_\xi(y)
    :=\widehat v_2(\xi,y)
      +\frac{1}{i\xi}\partial_y\widehat v_3(\xi,y),
    \qquad
    \mathbf g(\xi)
    :=\mathbf v_2(\xi)+\frac{1}{i\xi}D\mathbf v_3(\xi).
\end{equation*}
Thus \(\mathbf g(\xi)=(g_k(\xi))_{k\geq0}\) is precisely the Hermite
coefficient sequence of \(g_\xi\), and \eqref{eq:Born-factorization}
becomes
\begin{equation}
    \mathcal B^{23}_{m,p}(E)
    =\sum_{k\geq0}g_k(\xi)
      \int_{\mathbb R}A_2(y)\widetilde\varphi_k(y)\,dy.
    \label{eq:Born-reduced-v23}
\end{equation}
Consequently, the \(v_2,v_3\) contribution to every measurement at
frequency \(\xi\) depends only on \(\mathbf g(\xi)\).

We now rewrite all four contributions in Hermite coordinates. For a
coefficient sequence \(\mathbf a=(a_k)_{k\geq0}\), set
\begin{equation*}
    \mathcal T_{r,s}\mathbf a
    :=
    \sum_{k\geq0}
    \langle\varphi\rangle_{(r,s;k)}a_k=\sum_{k\ge 0}a_k
  \int \varphi_r(y) \varphi_s(y) \tilde{\varphi}_k(y) dy,\quad \forall\,r,s \in \N.
\end{equation*}
and use the convention \(\mathcal T_{r,s}\mathbf a:=0\) if
\(r<0\) or \(s<0\). The symmetry of the first two entries of the
triple Hermite tensor gives
\(\mathcal T_{r,s}=\mathcal T_{s,r}\). For a nonzero mode
\(\ell=(r,\epsilon_\ell)\), write
\begin{equation*}
    \phi_\ell(y;E)=c_\ell(E)\Phi_\ell(y;E),
    \qquad
    \Phi_\ell(y;E)
    :=\binom{\rho_\ell\varphi_{r-1}(y)}
                 {\tau_\ell(E)\varphi_r(y)},
\end{equation*}
where
$
    \rho_\ell:=\sqrt{2r},
$ and $
    \tau_\ell(E):=E-\xi_\ell(E).
$
For the exceptional mode, we adopt the conventions
\begin{equation}
    \Phi_{(0,-)}(y;E):=\binom{0}{\varphi_0(y)},
    \qquad
    \rho_{(0,-)}:=0,
    \qquad
    \tau_{(0,-)}:=1,
    \qquad
    c_{(0,-)}:=1.
    \label{eq:exceptional-unnormalized-spinor}
\end{equation}

Let \(m=(n,\epsilon_m)\), \(p=(q,\epsilon_p)\), and suppose that
\(\xi_{m,p}(E)=\xi\). Removing the known nonzero prefactor in
\eqref{eq:forward data} and the two spinor normalization constants,
define
\begin{equation}
    \mathfrak d_{m,p}(E)
    :=
    \frac{i\sqrt{|\xi_m(E)|\,|\xi_p(E)|}}
    {|E|\,c_m(E)c_p(E)}
    \widehat S^{\mathrm{lin}}_{m,p}(E).
    \label{eq:normalized-linearized-data-C}
\end{equation}
The dependence of \(\mathbf v_j=\mathbf v_j(\xi)\) and
\(\mathbf g=\mathbf g(\xi)\) on the fixed frequency will be suppressed
below. Direct multiplication of the spinor components, together with
\eqref{eq:Born-reduced-v23}, gives
\begin{align}
    \mathfrak d_{m,p}(E)
    ={}&
    \rho_m\rho_p\mathcal T_{n-1,q-1}\mathbf v_0
    +\tau_m\tau_p\mathcal T_{n,q}\mathbf v_0
    \nonumber\\
    &+\rho_m\tau_p\mathcal T_{n-1,q}\mathbf v_1
    +\tau_m\rho_p\mathcal T_{n,q-1}\mathbf v_1
    \nonumber\\
    &-i\rho_m\tau_p\mathcal T_{n-1,q}\mathbf g
    +i\tau_m\rho_p\mathcal T_{n,q-1}\mathbf g,
    \label{eq:reduced-Hermite-row}
\end{align}
where \(\tau_m=\tau_m(E)\) and \(\tau_p=\tau_p(E)\). The conventions
in \eqref{eq:exceptional-unnormalized-spinor} make this identity valid
when either mode is exceptional. Introduce the scalar moments
\begin{equation}
    M_r^{(0)}:=\mathcal T_{r,0}\mathbf v_0,
    \qquad
    M_r^{(1)}:=\mathcal T_{r,0}\mathbf v_1,
    \qquad
    M_r^{(g)}:=\mathcal T_{r,0}\mathbf g,
    \qquad r\in \N,
    \label{eq:Hermite-moments}
\end{equation}
and set all three types of moments equal to zero for negative indices.

We first use the two ordered pairs involving the exceptional mode. For
\(s\geq1\), set
\begin{equation*}
    E_s(\xi):=\frac{\xi}{2}+\frac{s}{\xi},
    \qquad
    \epsilon_s(\xi)
    :=\operatorname{sgn}\left(\frac{\xi^2-2s}{\xi}\right).
\end{equation*}
Introduce the two nonzero modes
\begin{equation*}
    m_s:=(s,\epsilon_s(\xi)),
    \qquad
    \widetilde m_s:=(s,-\epsilon_s(\xi)).
\end{equation*}
Lemma~\ref{lemma:Emp} gives
\begin{equation*}
    \xi_{m_s,(0,-)}\bigl(E_s(\xi)\bigr)
    =\xi_{(0,-),\widetilde m_s}\bigl(-E_s(\xi)\bigr)
    =\xi.
\end{equation*}
Moreover,
\begin{equation}
    \rho_{m_s}=\rho_{\widetilde m_s}=\sqrt{2s},
    \qquad
    \tau_{m_s}\bigl(E_s(\xi)\bigr)=\frac{2s}{\xi},
    \qquad
    \tau_{\widetilde m_s}\bigl(-E_s(\xi)\bigr)=-\frac{2s}{\xi}.
    \label{eq:zero-mode-spinor-coefficients}
\end{equation}
Define the corresponding normalized data by
\begin{equation*}
    \mathfrak d_s^{(1)}
    :=\mathfrak d_{m_s,(0,-)}\bigl(E_s(\xi)\bigr),
    \qquad
    \mathfrak d_s^{(2)}
    :=\mathfrak d_{(0,-),\widetilde m_s}
      \bigl(-E_s(\xi)\bigr).
\end{equation*}
Applying \eqref{eq:reduced-Hermite-row} to these two pairs yields
\begin{equation}    \label{eq:paired-zero-mode-rows}
    \begin{aligned}
    \mathfrak d_s^{(1)}
    &={}
    \frac{2s}{\xi}M_s^{(0)}
    +\sqrt{2s}\,M_{s-1}^{(1)}
    -i\sqrt{2s}\,M_{s-1}^{(g)},
\\
    \mathfrak d_s^{(2)}
    &={}
    -\frac{2s}{\xi}M_s^{(0)}
    +\sqrt{2s}\,M_{s-1}^{(1)}
    +i\sqrt{2s}\,M_{s-1}^{(g)}.
\end{aligned}
\end{equation}

Therefore
\begin{align}
    Y_s
    &:=\frac12\left(\mathfrak d_s^{(1)}
                    +\mathfrak d_s^{(2)}\right)
      =\sqrt{2s}\,M_{s-1}^{(1)},
    \label{eq:Ys-v1}\\
    Z_s
    &:=\frac12\left(\mathfrak d_s^{(1)}
                    -\mathfrak d_s^{(2)}\right)
      =\frac{2s}{\xi}M_s^{(0)}
       -i\sqrt{2s}\,M_{s-1}^{(g)}.
    \label{eq:Zs-v0-g}
\end{align}
In particular, the data determine every moment \(M_r^{(1)}\).

The map from Hermite coefficients to these moments is lower triangular.
Indeed,
\begin{equation}
    \langle\varphi\rangle_{(r,0;k)}=0\quad(k>r),
    \qquad
    \langle\varphi\rangle_{(r,0;r)}=2^{-r/2}\neq0.
    \label{eq:triangular-Hermite-moments}
\end{equation}
Consequently, the sequence \((M_r^{(1)})_{r\geq0}\) determines
\(\mathbf v_1\) by forward substitution. Notice, however, that each
\(Z_s\) contains one moment of \(\mathbf v_0\) and one moment of
\(\mathbf g\), so the two exceptional-mode measurements alone do not
separate these two unknowns.

To perform the remaining elimination, fix \(s\geq1\). By
Lemma~\ref{lemma:Emp}, we may choose propagating nonzero modes
$
    m=(s,\epsilon_m),
    \
    p=(1,\epsilon_p),
$
and an energy \(E\) such that \(\xi_m(E)-\xi_p(E)=\xi\). The dispersion relations imply
\begin{equation*}
    \tau_p-\tau_m=\xi,
    \qquad
    \tau_m(2E-\tau_m)=2s,
    \qquad
    \tau_p(2E-\tau_p)=2,
    \label{eq:tau-dispersion-identities}
\end{equation*}
and hence
\begin{equation}
    \xi=\frac{2s}{\tau_m}-\frac{2}{\tau_p},\qquad\frac{2}{\xi}\bigl(s\tau_p-\tau_m\bigr)
    =\tau_m\tau_p.
    \label{eq:key-tau-identity}
\end{equation}

Subtracting the now known \(\mathbf v_1\) contribution from the
normalized datum, define
\begin{equation}
    \mathfrak r_s
    :=\mathfrak d_{m,p}(E)
      -\rho_m\tau_p\mathcal T_{s-1,1}\mathbf v_1
      -\tau_m\rho_p\mathcal T_{s,0}\mathbf v_1.
    \label{eq:residual-data-s}
\end{equation}
Equation \eqref{eq:reduced-Hermite-row} then gives
\begin{align}
    \mathfrak r_s
    ={}&
    \rho_m\rho_p\mathcal T_{s-1,0}\mathbf v_0
    +\tau_m\tau_p\mathcal T_{s,1}\mathbf v_0
    \nonumber\\
    &-i\rho_m\tau_p\mathcal T_{s-1,1}\mathbf g
    +i\tau_m\rho_p\mathcal T_{s,0}\mathbf g.
    \label{eq:residual-Hermite-row}
\end{align}
The elementary Hermite recurrence gives, for any coefficient sequence
\(\mathbf a\),
\begin{align}
    \mathcal T_{s,1}\mathbf a
    &=\sqrt{s+1}\,\mathcal T_{s+1,0}\mathbf a
      +\sqrt{s}\,\mathcal T_{s-1,0}\mathbf a,
    \label{eq:T-s1-recurrence}\\
    \mathcal T_{s-1,1}\mathbf a
    &=\sqrt{s}\,\mathcal T_{s,0}\mathbf a
      +\sqrt{s-1}\,\mathcal T_{s-2,0}\mathbf a.
    \label{eq:T-sm1-1-recurrence}
\end{align}
Using~\eqref{eq:T-s1-recurrence} and~\eqref{eq:T-sm1-1-recurrence} in \eqref{eq:residual-Hermite-row}, we obtain
\begin{equation}    \label{eq:residual-before-elimination}
    \begin{aligned}
    \mathfrak r_s
    ={}&
    2\sqrt{s}\,M_{s-1}^{(0)}
    +\tau_m\tau_p
      \left(\sqrt{s+1}\,M_{s+1}^{(0)}
            +\sqrt{s}\,M_{s-1}^{(0)}\right)
\\
    &-i\sqrt{2s}\,\tau_p
      \left(\sqrt{s}\,M_s^{(g)}
            +\sqrt{s-1}\,M_{s-2}^{(g)}\right)
      +i\sqrt2\,\tau_m M_s^{(g)}.
\end{aligned}
\end{equation}

Set \(Z_0:=0\), consistently with the convention on negative moments.
Equation \eqref{eq:Zs-v0-g} at the indices \(s+1\) and \(s-1\) reads
\begin{equation}    \label{eq:neighboring-Z-identities}
    \begin{aligned}
    Z_{s+1}
    &=\frac{2(s+1)}{\xi}M_{s+1}^{(0)}
      -i\sqrt{2(s+1)}\,M_s^{(g)},
\\
    Z_{s-1}
    &=\frac{2(s-1)}{\xi}M_{s-1}^{(0)}
      -i\sqrt{2(s-1)}\,M_{s-2}^{(g)}.
\end{aligned}
\end{equation}

For \(s=1\), the second line is understood as \(Z_0=0\). Substituting
these expressions into \eqref{eq:residual-before-elimination} and using
\eqref{eq:key-tau-identity} cancels the terms containing
\(M_{s+1}^{(0)}\), \(M_s^{(g)}\), and \(M_{s-2}^{(g)}\), leaving
\begin{equation}
    M_{s-1}^{(0)}
    =\frac{1}{4\sqrt{s}}
      \left(
        \mathfrak r_s
        -\frac{s\tau_p-\tau_m}{\sqrt{s+1}}Z_{s+1}
        -\sqrt{s}\,\tau_p Z_{s-1}
      \right),
    \qquad s\geq1.
    \label{eq:recover-v0-moments}
\end{equation}
Thus all the moments \(M_r^{(0)}\) are determined. Equation
\eqref{eq:Zs-v0-g} then gives
\begin{equation}
    M_{s-1}^{(g)}
    =\frac{\dfrac{2s}{\xi}M_s^{(0)}-Z_s}
           {i\sqrt{2s}},
    \qquad s\geq1,
    \label{eq:recover-g-moments}
\end{equation}
so all the moments \(M_r^{(g)}\) are determined as well. Applying the
same lower-triangular inversion \eqref{eq:triangular-Hermite-moments}
to \((M_r^{(0)})_{r\geq0}\) and \((M_r^{(g)})_{r\geq0}\) recovers
\(\mathbf v_0\) and \(\mathbf g\), respectively. Together with the
recovery of \(\mathbf v_1\), this proves the asserted uniqueness of
\[
    \widehat v_0(\xi,\cdot),
    \qquad
    \widehat v_1(\xi,\cdot),
    \qquad
    g_\xi
    =\widehat v_2(\xi,\cdot)
      +\frac{1}{i\xi}\partial_y\widehat v_3(\xi,\cdot).
\]

It remains to describe precisely the ambiguity in \(v_2,v_3\). For two
potentials with the same data, let \(\delta\mathbf v_j\) denote the
difference of their Hermite coefficient sequences. The uniqueness just
proved implies
\[
    \delta\mathbf v_0=\delta\mathbf v_1=0,
    \qquad
    \delta\mathbf v_2
      +\frac{1}{i\xi}D\delta\mathbf v_3=0.
\]
Since \(\xi\neq0\), the last identity is equivalent to
\begin{equation}
    (\delta\mathbf v_2,\delta\mathbf v_3)
    =(-D\mathbf h,i\xi\mathbf h),
    \qquad
    \mathbf h:=\frac{1}{i\xi}\delta\mathbf v_3.
    \label{eq:exact-Hermite-kernel}
\end{equation}
Conversely, every \(\mathbf h\) for which the pair on the right-hand side
belongs to the admissible coefficient class leaves \(\mathbf g\), and
hence all the fixed-frequency linearized data, unchanged. This is the
exact coefficient-level kernel.

If
\[
    \widehat\theta(\xi,y)
    :=\sum_{k\geq0}h_k(\xi)\widetilde\varphi_k(y),
\]
then \eqref{eq:exact-Hermite-kernel} becomes
\begin{equation}
    \delta\widehat v_2(\xi,y)
    =-\partial_y\widehat\theta(\xi,y),
    \qquad
    \delta\widehat v_3(\xi,y)
    =i\xi\widehat\theta(\xi,y),
    \label{eq:frequencywise-gauge-kernel}
\end{equation}
which is the Fourier--Hermite form of
\[
    \delta V
    =-(\partial_y\theta)\sigma_2
      +(\partial_x\theta)\sigma_3.
\]
Finally, with the Fourier convention used in the paper,
\begin{equation}
    i\xi g_\xi
    =i\xi\widehat v_2(\xi,\cdot)
      +\partial_y\widehat v_3(\xi,\cdot)
    =\reallywidehat{\partial_xv_2+\partial_yv_3}(\xi,\cdot).
    \label{eq:recoverable-divergence}
\end{equation}
Thus the fixed-frequency data determine the Fourier component at \(\xi\)
of \(\partial_xv_2+\partial_yv_3\), completing the proof.
\end{proof}

\begin{remark}\label{rem:frequencywise-versus-global-gauge}
The kernel description \eqref{eq:exact-Hermite-kernel} is a statement at
each fixed longitudinal frequency. Every admissible real-valued gauge
function \(\theta\) produces a family of the form
\eqref{eq:frequencywise-gauge-kernel}, in agreement with
Proposition~\ref{prop:compact-gauge}. The converse should not be inferred
from a single frequency, however: an arbitrary frequency-dependent family
\(\mathbf h(\xi)\) need not assemble into the Fourier transform of one
real-valued, sufficiently regular, compactly supported function \(\theta\).
Such a global conclusion additionally requires the appropriate compatibility,
regularity, and support conditions in \(\xi\). What follows directly from
the reconstruction above is that data over all admissible frequencies
determine \(\partial_xv_2+\partial_yv_3\) by Fourier inversion, whereas the
gauge transformation in Proposition~\ref{prop:compact-gauge} gives a global
family of perturbations invisible to the scattering data.
\end{remark}

\section{Algorithm for the computation of TR matrices in the Dirac model}\label{app:TRalg}
%%%

When solving~\eqref{eq:outgoing solution}, if the support of $V$ is large, 
accurate computation requires a high-order quadrature rule in the $x$-direction. 
To avoid this,~\cite{bal2023asymmetric} proposed to decompose the domain into 
small subintervals in the $x$-direction, compute the corresponding TR matrices for each subinterval, and then iteratively merge the TR matrices of adjacent intervals 
until the TR matrix for the entire domain is obtained.

\paragraph{Merging two TR Matrices}
We now list the merging formula for two adjacent intervals. 
Let $L$ and $R$ denote the TR matrices associated with two neighboring intervals 
$I_L$ and $I_R$, respectively. 
Then, the TR matrix for the combined interval $I_L \cup I_R$ is given by
\begin{equation}\label{scattermatrix_formula}
\begin{pmatrix}
L_{11}(I - R_{12}L_{21})^{-1}R_{11} 
& 
L_{11}(I - R_{12}L_{21})^{-1}R_{12}L_{22} + L_{12} \\[1ex]
R_{22}(I - L_{21}R_{12})^{-1}L_{21}R_{11} + R_{21} 
& 
R_{22}(I - L_{21}R_{12})^{-1}L_{22}
\end{pmatrix}.
\end{equation}
\begin{algorithm}[ht!]
	\caption{Computing density and eigenfunction in a single slab (leaf)}  \label{alg:single slab}
	\begin{algorithmic}[1]
	\Require Potential Field $V$; Interval of $V$ that is compactly supported $I=[x_L,x_R]$; Level of binary merging $L$;  Incoming wave condition $\alpha_+(x_L)$ and $\alpha_-(x_R)$; Discretization configuration $(n_x,n_y)$.
	\Ensure Eigenfunction $\hat{\psi}$ given incoming condition; Outgoing wave coefficients $\alpha_-(x_L)$ and $\alpha_+(x_R)$.
	\State Construct $\hat{V}$ and $\hat{G}$ projected to $n_x$ Legendre polynomials and $n_y$ Hermite functions. 
	\State Compute $\Pi_{\mathcal{V}}\psi_{in}$ with $\alpha_-(x_L)$ and $\alpha_+(x_R)$.
	\State Solve $\hat{\rho}$ by $\hat{\rho}=-(I+\hat{V}\hat{G})^{-1}\hat{V}\Pi_\mathcal{V} \psi_{in}$.
	
	\State Recover $\psi$ by $\hat{\psi}(x,y)=\psi_{in}(x,y)+\int G(x,y;x_0,y_0)\hat{\rho}(x_0,y_0)dx_0dy_0$.
    \State Extract $\alpha_{n-}(x_L)=e^{i\xi_{n-}(x_L-x_R)}\alpha_{n-}(x_R)+\int \overline{\vartheta_{n,-}(y)}G(x_L,y;x_0,y_0)\hat{\rho}(x_0,y_0)dx_0dy_0dy$.
    \State Extract $\alpha_{n+}(x_R)=e^{i\xi_{n+}(x_R-x_L)}\alpha_{n+}(x_L)+\int \overline{\vartheta_{n,+}(y)}G(x_R,y;x_0,y_0)\hat{\rho}(x_0,y_0)dx_0dy_0dy$.
	\end{algorithmic}
\end{algorithm}
\begin{algorithm}[ht!]
	\caption{Merging Algorithm}  
    \label{alg:eigenfunctions in a slab}
    	\begin{algorithmic}[1]
	\Require Potential Field $V$; Interval $I=[x_L,x_R]$ which support $V$; Level of binary merging $L$; (optional) Incoming wave condition $\alpha_+(x_L)$ and $\alpha_-(x_R)$.
	\Ensure TR matrix $M_{2^L}$ of interval $I$; (optional) eigenfunctions $\hat{\psi}$ given incoming wave $\alpha_+(x_L)$ and $\alpha_-(x_R)$. 
	\State Partition $I$ to be $2^L$ intervals. Denote the intervals as $\{I_i\}$, $i=1,2,\cdots, 2^L$ and the grid points as $x_i$, $i=0,1,\cdots, 2^L$.
	\For{$i$ in $1,\cdots ,2^L$}
	\State Compute TR matrix $T_{i}$ with potential field $V$ limited on interval $I_i$ by Alg.\ref{alg:single slab}.
	\EndFor
	\State Let $M_{1}=T_{1}$.
	\For{$i$ in $1\rightarrow2^L-1$}
        \State  Use equation \eqref{scattermatrix_formula} to calculate the TR Matrix $M_{i+1}$ for $[x_0,x_{i+1}]$ by merging $T_{i+1}$ and $M_i$.
        \State Use equation \eqref{eq:midmatrix} to Calculate the Intersection coefficient matrix $\mathcal{M}_{i}$ with intersection point $x_i$ on interval $[0,x_{i+1}]$  by merging $M_{i}$ with $T_{i+1}$.
	\EndFor \\
    (optional)
	\State Assign $\alpha_+(x_0)$ with $\alpha_+(x_L)$ and $\alpha_-(x_{2^L})$ with $\alpha_-(x_R)$.
        \For{$i$ in $2^L-1\rightarrow 1$}
        \State $\begin{pmatrix}
            \alpha_{-}(x_i)\\ \alpha_{+}(x_i)
        \end{pmatrix}=
        \mathcal{M}_{i}\begin{pmatrix}
            \alpha_{-}(x_{i+1})\\ \alpha_{+}(x_0)
        \end{pmatrix}$
        \EndFor
        \For{$i$ in $1\rightarrow2^L$}
        \State Use the coefficients $\alpha_{n-}(x_i)$ and $\alpha_{n+}(x_{i-1})$ as input for Algorithm \ref{alg:single slab} to recover $\hat{\psi}$ on interval $I_i$.
	\EndFor
        \end{algorithmic}
\end{algorithm}
After computing the TR matrix on the whole interval, to reconstruct the coefficients $\alpha$ inside the interval, we can use the intersection coefficient matrix $\mathcal{M}$, defined as,
\begin{equation}\label{eq:midmatrix}
\begin{pmatrix}
\alpha_{-,\mathcal{M}} \\[2pt]
\alpha_{+,\mathcal{M}}
\end{pmatrix}
=
\mathcal{M}
\begin{pmatrix}
\alpha_{-,R} \\[2pt]
\alpha_{+,L}
\end{pmatrix},
\end{equation}
where 
$\begin{pmatrix}\alpha_{-,\mathcal{M}} \\ \alpha_{+,\mathcal{M}}\end{pmatrix}$ 
denotes the Fourier coefficients projected onto the unperturbed eigenfunction basis 
$\{\phi_m\}$ at the intersection point. 
The matrix $\mathcal{M}$ can be computed explicitly as
\begin{equation}\label{eq:computing mid matrix}
\mathcal{M} =
\begin{pmatrix}
(I - R_{12}L_{21})^{-1}R_{11} 
& 
(I - R_{12}L_{21})^{-1}R_{12}L_{22} \\[1ex]
(I - L_{21}R_{12})^{-1}L_{21}R_{11} 
& 
(I - L_{21}R_{12})^{-1}L_{22}
\end{pmatrix}.
\end{equation}

In what follows, for completeness, we list the Green's function approach to compute eigenfunctions of a perturbed Dirac operator in Alg.\ref{alg:single slab} and the merging algorithms based on TR matrix operations in Alg.\ref{alg:eigenfunctions in a slab}. It is noted that, throughout all numerical experiments reported in this work,
the merging level is fixed to $L=0$, so that no recursive merging is performed and
the TR matrices are computed directly on a single interval.

%%%%
%\newpage
\bibliographystyle{plain}
\bibliography{ref,ref2}
\end{document}